\documentclass[aps,prx,twocolumn,groupedaddress,superscriptaddress,longbibliography,nofootinbib]{revtex4-1}

\usepackage{amsmath,amssymb}
\usepackage{graphicx}
\usepackage{multirow}
\usepackage{bm}
\usepackage{mathtools}
\usepackage{dsfont}
\usepackage{amsfonts}
\usepackage{xcolor}
\usepackage[normalem]{ulem}
\usepackage{url}
\usepackage{wasysym}

\usepackage[colorlinks=true,linkcolor=magenta,citecolor=magenta,hypertexnames=false]{hyperref}

\def\ba#1\ea{\begin{align} #1\end{align}}
\def\bg#1\eg{\begin{gather}#1\end{gather}}
\def\bpm{\begin{pmatrix}}
\def\epm{\end{pmatrix}}
\newcommand{\nn}{\nonumber}

\renewcommand{\b}[1]{{\boldsymbol #1}}
\newcommand{\bx}{\b x}
\newcommand{\bk}{\b k}

\newcommand{\mc}[1]{\mathcal{#1}}
\newcommand{\mf}[1]{\mathfrak{#1}}
\newcommand{\der}{\partial}
\newcommand{\dg}{\dagger}
\newcommand{\om}{\omega}
\newcommand{\sg}{\sigma}
\newcommand{\vph}{\varphi}

\newcommand{\ep}{\epsilon}

\newcommand{\ket}[1]{| #1 \rangle}
\newcommand{\bra}[1]{\langle #1 |}
\newcommand{\brk}[2]{\langle #1 | #2 \rangle}

\newcommand{\eq}[1]{Eq.~\eqref{#1}}
\newcommand{\fig}[1]{Fig.~\ref{#1}}
\newcommand{\figs}[1]{Figs.~\ref{#1}}

\newcommand{\dm}{d_{\rm max}}
\newcommand{\ra}{\rightarrow}
\newcommand{\rbs}{r_{\rm BS}}
\newcommand{\ourtitle}{Wave-function geometry of band crossing points in two-dimensions}

\newcommand{\magenta}[1]{\textcolor{magenta}{#1}}

\begin{document}
\title{\ourtitle}

\author{Yoonseok \surname{Hwang}}
\thanks{These authors contributed equally.}
\affiliation{Center for Correlated Electron Systems, Institute for Basic Science (IBS), Seoul 08826, Korea}
\affiliation{Department of Physics and Astronomy, Seoul National University, Seoul 08826, Korea}
\affiliation{Center for Theoretical Physics (CTP), Seoul National University, Seoul 08826, Korea}

\author{Junseo \surname{Jung}}
\thanks{These authors contributed equally.}
\affiliation{Center for Correlated Electron Systems, Institute for Basic Science (IBS), Seoul 08826, Korea}
\affiliation{Department of Physics and Astronomy, Seoul National University, Seoul 08826, Korea}

\author{Jun-Won \surname{Rhim}}
\affiliation{Center for Correlated Electron Systems, Institute for Basic Science (IBS), Seoul 08826, Korea}
\affiliation{Department of Physics and Astronomy, Seoul National University, Seoul 08826, Korea}
\affiliation{Department of Physics, Ajou University, Suwon 16499, Korea}

\author{Bohm-Jung \surname{Yang}}
\email{bjyang@snu.ac.kr}
\affiliation{Center for Correlated Electron Systems, Institute for Basic Science (IBS), Seoul 08826, Korea}
\affiliation{Department of Physics and Astronomy, Seoul National University, Seoul 08826, Korea}
\affiliation{Center for Theoretical Physics (CTP), Seoul National University, Seoul 08826, Korea}

\begin{abstract}
Geometry of the wave function is a central pillar of modern solid state physics.
In this work, we unveil the wave-function geometry of two-dimensional semimetals with band crossing points (BCPs).
We show that the Berry phase of BCPs are governed by the quantum metric describing the infinitesimal distance between quantum states.
For generic linear BCPs, we show that the corresponding Berry phase is determined either by an angular integral of the quantum metric, or equivalently, by the maximum quantum distance of Bloch states.
This naturally explains the origin of the $\pi$-Berry phase of a linear BCP.
In the case of quadratic BCPs, the Berry phase can take an arbitrary value between 0 and $2\pi$.
We find simple relations between the Berry phase, maximum quantum distance, and the quantum metric in two cases:
(i) when one of the two crossing bands is flat; (ii) when the system has rotation and/or time-reversal symmetries.
To demonstrate the implication of the continuum model analysis in lattice systems, we study tight-binding Hamiltonians describing quadratic BCPs.
We show that, when the Berry curvature is absent, a  quadratic BCP with an arbitrary Berry phase always accompanies another quadratic BCP so that 
the total Berry phase of the periodic system becomes zero.
This work demonstrates that the quantum metric plays a critical role in understanding the geometric properties of topological semimetals.
\end{abstract}

\maketitle

{\it \magenta{Introduction.|}}
The Berry phase of electronic wave functions can have profound effects on vast physical phenomena in condensed matter~\cite{berry1984quantal,zak1989berry,vanderbilt1993electric,xiao2010berry}.
The significance of the Berry phase lies in the fact that it is not only gauge-invariant (up to an integer multiple of $2\pi$), but also geometric.
For instance, the Berry phase, normally written as a line integral of the Berry connection over a loop in the parameter space, can also be expressed as a surface integral of the Berry curvature so that it can be understood as an Aharonov-Bohm phase arising from the Berry gauge flux.
The geometric interpretation of the Berry phase in terms of the Berry curvature answers the origin of the anomalous Hall effect~\cite{ong2006geometry} and also allows us to include various topological phenomena in the realm of the Berry-phase-related physics~\cite{xiao2010berry}.

Interestingly, recent studies of topological phases have shown that the Berry phase can also serve as a topological invariant~\cite{hasan2010colloquium,qi2011topological,chiu2016classification}.
For instance, in a class of topological semimetals having band crossing nodes, the stability of a nodal point in two dimensions or a nodal line in three dimensions is guaranteed by the quantized $\pi$-Berry phase defined along a loop enclosing the node in momentum space.
However, when applied to such band crossing points (BCPs), the geometric interpretation of the Berry phase in terms of the Berry curvature does not work unless a singular source of Berry curvature is introduced.
This is because the presence of a band degeneracy inside the loop, on which the Berry phase is defined, prohibits transforming the line integral for the Berry phase to the surface integral with the Berry curvature. 
In fact, the quantization of Berry phase requires the Berry curvature to vanish because, otherwise, the Berry phase for a BCP becomes path dependent.
This indicates that the geometric character of the Berry phase describing BCPs should have distinct nature, independent of the Berry curvature.

In this work, we unveil the wave-function geometry of BCPs in two-dimensional (2D) crystals.
Explicitly, we show that the Berry phase is completely determined by the quantum metric~\cite{provost1980riemannian,berry1989quantum,anandan1990geometry,zanardi2007information,cheng2010quantum,resta2011insulating,kolodrubetz2017geometry}, which describes the infinitesimal distance between two wave functions in the parameter space.
Together with the Berry curvature, the quantum metric constitutes the quantum geometric tensor, which fully characterizes the geometry of quantum states.
We first show that the maximum quantum distance between the Bloch states around a linear BCP (LBCP) takes the largest allowed value 1 as determined by an angular integral of the quantum metric along a loop enclosing the LBCP.
This characteristic property of LBCPs gives rise to the quantized value $\pi$ of the Berry phase.

In the case of quadratic BCPs (QBCPs)~\cite{chong2008effective,sun2009topological}, we show that the path-independent Berry phase can take an arbitrary value depending on the Hamiltonian parameters, which modify the quantum metric distribution.  
We find simple relations between the geometric quantities characterizing the BCPs such as the Berry phase, quantum metric, and maximum quantum distance, in two cases.
One is when one of the two crossing bands is flat~\cite{bergman2008band,dora2014occurrence,rhim2019classification,ma2020spin,rhim2020quantum,ma2020direct,rhim2021singular}.
The other is when the system has rotation or time-reversal symmetries.
In both cases, we find that the Berry phase of a QBCP is determined by an angular integral of the quantum metric along a loop enclosing it, which is proportional to the maximum quantum distance of relevant Bloch states.

To demonstrate the implication of the continuum model analysis for the periodic lattice systems, we study tight-binding models describing QBCPs.
In the case with vanishing Berry curvature over the whole Brillouin zone (BZ), we find that a QBCP with an arbitrary value of Berry phase always accompanies another BCP.
On the other hand, when the Berry curvature is finite, we show that a single QBCP with an arbitrary Berry phase can exist in the BZ.
In both lattice models, the obtained geometric quantities of QBCPs are consistent with our continuum theory.

{\it \magenta{$\pi$-Berry phase of LBCPs.|}}
The quantized $\pi$-Berry phase of an LBCP (or a Dirac point)~\cite{novoselov2005two,zhang2005experimental,hasan2010colloquium} has been understood as follows.
For a given LBCP, its Berry phase is determined by the line integral of the Berry connection along a loop $\ell$ enclosing it in momentum space.
According to Stokes theorem, the difference of the Berry phases computed along two different loops $\ell_1,~\ell_2$ enclosing the LBCP is given by the integral of the Berry curvature over the area $S_{\ell_1,\ell_2}$ bounded by $\ell_1,~\ell_2$.
Then the Berry phase can be path independent only when the Berry curvature integral over $S_{\ell_1,\ell_2}$ vanishes for any choice of $\ell_1,~\ell_2$.

Normally, the Berry curvature integral vanishes when suitable symmetry exists such as space-time inversion~\cite{fang2015new,ahn2017unconventional} or mirror symmetries~\cite{fang2015new} (see Supplemental Materials~\cite{supple}).
Below we show that the Berry phase quantization of LBCPs does not rely on the symmetry, but originates from the peculiar geometry of Dirac spinors.
The main role of symmetry is to forbid mass terms so that symmetry-protected LBCPs can form a stable Dirac semimetal phase.
Even an unstable LBCP appearing at the critical point between insulators has $\pi$-Berry phase.

{\it \magenta{Quantum distance and quantum metric.|}}
To describe the quantum geometry of BCPs, we define several geometric concepts.
The Hilbert-Schmidt quantum distance~\cite{provost1980riemannian,buvzek1996quantum,witte1999new,dodonov2000hilbert} between two states $\ket{\psi(\bk)}$ and $\ket{\psi(\bk^\prime)}$ is defined as
\ba
d^2(\bk,\bk^\prime) = 1-|\brk{\psi(\bk)}{\psi(\bk^\prime)}|^2,
\label{eq:dis_def}
\ea
which takes the maximal value 1 (minimal value 0) for two orthogonal (identical) states.
For two infinitesimally close states at the momentum $\bk$ and $\bk^\prime=\bk+d\bk$, respectively, 
\ba
d^2(\bk,\bk+d\bk)=\mf{G}_{ij}(\bk) dk_i dk_j,
\label{eq:qm_def_expansion}
\ea
where the quantum geometric tensor $\mf{G}_{ij}(\bk)$, which is Hermitian and gauge-invariant, is given by
\ba
\mf{G}_{ij}(\bk)
= \brk{\der_i \psi(\bk)}{\der_j \psi(\bk)} - A_i(\bk) A_j(\bk),
\label{eq:qgt_def}
\ea
in which $A_i(\bk)=i \brk{\psi(\bk)}{\der_i \psi(\bk)}$ indicates the Berry connection.
The real and imaginary parts of $\mf{G}_{ij}(\bk)$ correspond to the quantum metric $g_{ij}(\bk)$ and the Berry curvature $F_{ij}(\bk)$, respectively.

{\it \magenta{Two-band Hamiltonian and Bloch sphere.|}}
In general, BCPs between two non-degenerate bands can be described by a two-band Hamiltonian 
\bg
H(\bk) = f_0(\bk) \sg_0 - \b f(\bk) \cdot \b \sg,
\label{eq:H_twoband}
\eg
where $\b \sg=(\sg_1,\sg_2,\sg_3)$ denote the Pauli matrices, $\sg_0$ indicates a $2\times 2$ identity matrix, and $[f_0(\bk), \b f(\bk)]=[f_0(\bk),f_1(\bk),f_2(\bk),f_3(\bk)]$ are real functions of $\bk$.
The occupied state $\ket{\psi(\bk)}$ satisfies
\ba
[\check{\b n}(\bk) \cdot \b \sg]\ket{\psi(\bk)}=\frac{1}{2}\ket{\psi(\bk)},
\ea
and the corresponding energy eigenvalue is $f_0(\bk)-|\b f(\bk)|$.
Here, $\check{\b n}(\bk)$ denotes a point on the Bloch sphere $S^2_{\rm BS}$ with radius $\rbs=\frac{1}{2}$ defined by
\ba
\check{\b n}(\bk)=\frac{1}{2}\frac{\b f(\bk)}{|\b f(\bk)|} \in S^2_{\rm BS}.
\label{eq:nvec_def}
\ea

\begin{figure*}[t!]
\centering
\includegraphics[width=0.98\textwidth]{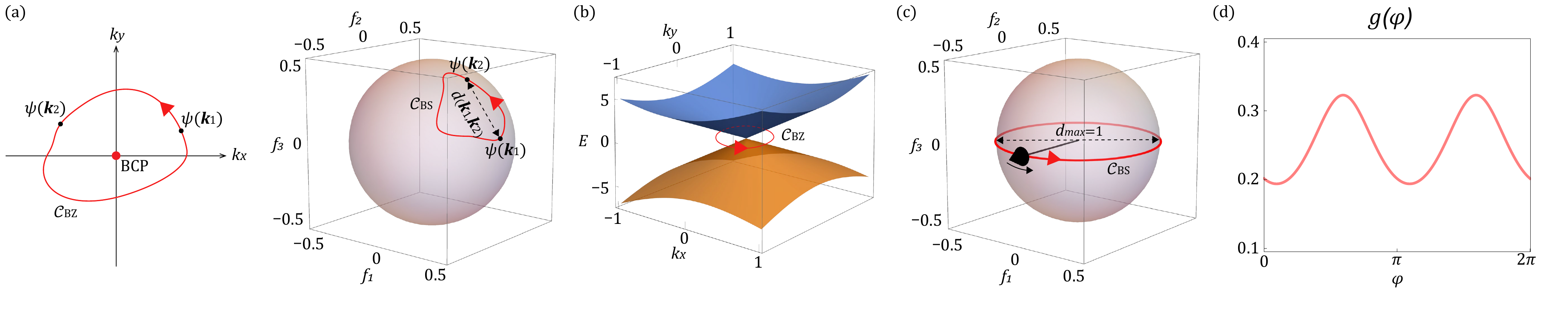}
\caption{
Wave-function geometry of LBCPs.
(a) Mapping between a closed path $\mc{C}_{\rm BZ}$ in momentum space (left) and the loop $\mc{C}_{\rm BS}$ on the Bloch sphere $S^2_{\rm BS}$ (right).
The straight-line distance between $\check{\b n}(\bk_{1,2})$ determines the quantum distance $d(\bk_1,\bk_2)$ between $\ket{\psi(\bk_{1,2})}$.
(b) The band structure around an LBCP at $\bk=(0,0)$ obtained by $H_{\rm L}(\bk)$ with $(t_1,t_2,t_3,b_1,b_2)=(3.9,0.25,-3.5,0,0)$.
(c) $\mc{C}_{\rm BS}$ on $S^2_{\rm BS}$ corresponding to $\mc{C}_{\rm BZ}$ in (b).
For an LBCP, the relevant $\mc{C}_{\rm BS}$ always forms a great circle, thus the maximum quantum distance is always $\dm=1$.
The red arrows on $\mc{C}_{\rm BZ}$ and $\mc{C}_{\rm BS}$ denote their orientation.
The big black arrow denotes $\check{\b n}(0,0)$ which moves counterclockwise as the momentum changes along $\mc{C}_{\rm BZ}$.
(d) Quantum metric $g(\phi)$ in the polar coordinates.
While the Berry curvature vanishes everywhere (except at the BCP), the quantum metric is generally non-zero.
The integral $\oint d\phi \sqrt{g(\phi)}=\pi \dm$ gives the quantized Berry phase $\Phi_B(\mc{C}_{\rm BZ})=\pi$ and $\dm=1$.
}
\label{Fig1}
\end{figure*}

From \eq{eq:nvec_def}, one can find several important relations between $\ket{\psi(\bk)}$ in the Hilbert space 
and $\check{\b n}(\bk)$ on $S_{\rm BS}^2$~\cite{supple}.
For this, let us consider a closed path $\mc{C}_{\rm BZ}$ enclosing the BCP in momentum space. Then another closed path $C_{\rm BS}$ on $S^2_{\rm BS}$ corresponding to $\mc{C}_{\rm BZ}$ is determined by \eq{eq:nvec_def} [see \fig{Fig1}(a)].
First, the quantum distance between $\ket{\psi(\bk_{1})}$ and $\ket{\psi(\bk_{2})}$ is equal to the straight-line distance between $\check{\b n}(\bk_{1})$ and $\check{\b n}(\bk_{2})$ on $S^2_{\rm BS}$ 
\ba
d(\bk_1,\bk_2) = |\check{\b n}(\bk_1)-\check{\b n}(\bk_2)|.
\label{eq:dis_BS}
\ea
We define the maximum quantum distance $\dm$ as the maximum value of $d(\bk_1,\bk_2)$ for $\bk_{1,2} \in \mc{C}_{\rm BZ}$.

Second, the length of $\mc{C}_{\rm BS}$ is given by an integration of the quantum metric along $\mc{C}_{\rm BZ}$:
\ba
\left| \mc{C}_{\rm BS} \right|
= \oint_{\mc{C}_{\rm BZ}} \sqrt{g_{ij}(\bk) dk_i dk_j}.
\label{eq:length}
\ea

Third, the Berry phase $\Phi_B(\mc{C}_{\rm BZ})=\oint_{\mc{C}_{\rm BZ}} d\bk\cdot \b A(\bk)$, defined along $\mc{C}_{\rm BZ}$, can also be described geometrically on $S^2_{\rm BS}$.
Then, $\Phi_B(\mc{C}_{\rm BZ})$ is given by a half of the solid angle $\Omega(\mc{C}_{\rm BS})$ on $S^2_{\rm BS}$ as
\ba
\Phi_B(\mc{C}_{\rm BZ}) = -\frac{1}{2}\Omega(\mc{C}_{\rm BS}), \quad ({\rm mod \,} 2\pi).
\label{eq:Berry_SolidAngle}
\ea

In general, there is no closed-form expression relating the three geometric quantities in Eqs.~\eqref{eq:dis_BS} to \eqref{eq:Berry_SolidAngle}.
However, for LBCPs and QBCPs, we demonstrate below the explicit formulas connecting them under the condition that $C_{\rm BS}$ becomes a circle.

{\it \magenta{Wave function geometry of LBCPs.|}}
The most general form of the $\bk$ linear Hamiltonian is 
\ba
H_{\rm L}^{(0)}(\bk) = (b_1 k_x + b_2 k_y) \sg_0 + \sum_{a=1}^{3} (v_{ax} k_x + v_{ay} k_y) \sg_a,
\label{eq:generic_linear}
\ea
where $b_{1,2}$, $v_{ai}$ ($a=1,2,3$, $i=x,y$) are constants.
In general, $H_{\rm L}^{(0)}(\bk)$ does not have any symmetry.
But its Berry curvature vanishes at every $\bk$ so that the BCP at $\bk=0$ has a path-independent Berry phase, either 0 or $\pi$, depending on $b_{1,2}$, $v_{ai}$.
Here, the Berry curvature vanishes because every term in $H_{\rm L}^{(0)}(\bk)$ is linear in $\bk$, \textit{i.e.}, $H_{\rm L}^{(0)}(\bk)$ is a homogeneous-order Hamiltonian of degree 1.

After successive unitary transformations~\cite{supple}, $H_{\rm L}^{(0)}(\bk)$ becomes
\ba
H_{\rm L}(\bk) =& t_3 k_y \sg_1 + (t_1 k_x + t_2 k_y) \sg_2 + (b_1 k_x + b_2 k_y) \sg_0.
\label{eq:linear_H}
\ea
We note that $t_{1,3}\ne0$, because otherwise, $H_{\rm L}^{(0)}(\bk)$ describes a nodal line, not a single LBCP.
Comparing \eq{eq:linear_H} to \eq{eq:H_twoband}, we find $\b f(\bk)=(-t_3 k_y,-t_1 k_x-t_2 k_y,0)$, which forms a plane passing through the origin in the three-dimensional space spanned by $[f_1(\bk),f_2(\bk),f_3(\bk)]$ when $\bk$ is varied.
For a closed path $\mc{C}_{\rm BZ}$ enclosing the origin [see \fig{Fig1}(b)], the corresponding $\mc{C}_{\rm BS}$ forms a great circle on $S_{\rm BS}^2$. 
In this case, the maximum quantum distance $\dm$ becomes 1.
Hence, we obtain a nodal point at $\bk=0$ with $\dm=1$.
The band structure, the Bloch sphere, and the quantum metric of $H_{\rm linear}(\bk)$ are shown in \figs{Fig1}(b) to \ref{Fig1}(d).

For an LBCP described by $H_{\rm L}(\bk)$, the quantum metric tensor $g_{\phi\phi}(\phi)\equiv g(\phi)$ takes a closed form~\cite{supple} which is plotted in \fig{Fig1}(d) as a function of $\phi=\tan^{-1} (k_y/k_x)$. 
As the eigenstates of $H_{\rm L}(\bk)$ are independent of $|\bk|$, they depend only on $\phi$ so that the relevant Berry curvature vanishes, which is generally valid for any homogeneous-order Hamiltonian.
As the Berry curvature is zero, the quantum metric is the only gauge-invariant geometric tensor.
When $\phi$ changes by $2\pi$ along a loop $\mc{C}_{\rm BZ}$, $\check{\b n}(\bk)$ also forms a closed loop $\mc{C}_{\rm BS}$ with the length $|\mc{C}_{\rm BS}|=\oint d\phi \sqrt{g(\phi)}=\pi$ [see \eq{eq:length}].
We note that $|\mc{C}_{\rm BS}|$ is also given by $\pi \dm$ since $\mc{C}_{\rm BS}$ is the great circle with diameter $\dm$.
The relevant Berry phase is $\Phi_B(\mc{C}_{\rm BZ})=\pi$, a half of the solid angle $\Omega(\mc{C}_{\rm BS})=2\pi$ as noted above [see \fig{Fig1}(c)].
The geometrical property of an LBCP can be summarized as follows:
\ba
\dm = \frac{1}{\pi} \oint d\phi \sqrt{g(\phi)} = 1, \quad \Phi_B(\mc{C}_{\rm BZ}) = \pi.
\label{eq:linear_sum}
\ea

\begin{figure*}[t!]
\centering
\includegraphics[width=0.98\textwidth]{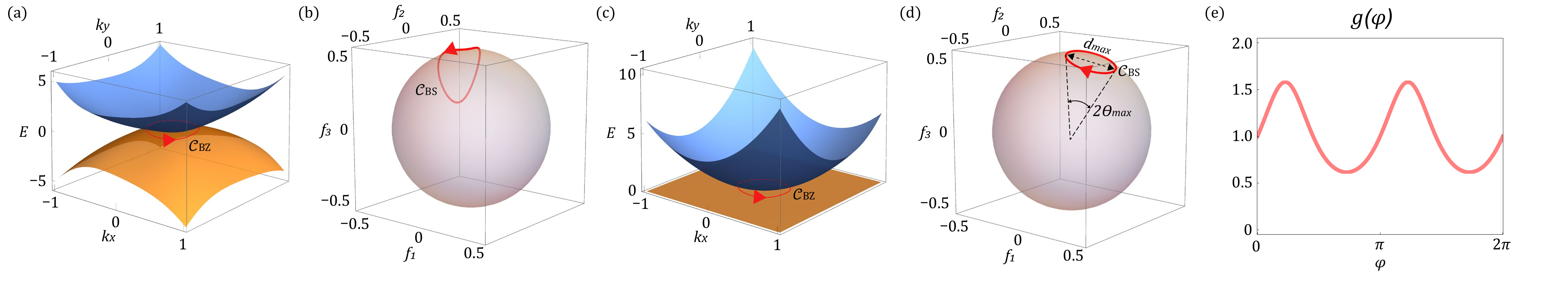}
\caption{
Wave-function geometry of QBCPs.
(a,b) The band structure and the relevant Bloch sphere around a QBCP at $\bk=(0,0)$ obtained from $H_{\rm Q}(\bk)$
with $(t_1,t_2,t_3,t_4,t_5,t_6,b_1,b_2,b_3)=(-3.8,-0.25,-0.35,1.8,-1.8,2.2,0,0,0)$.
$\mc{C}_{\rm BS}$ generally has an elliptical shape.
(c)-(e) The band structure, the Bloch sphere, and the quantum metric of a QBCP when one of the two crossing bands is flat, obtained from $H_{\rm flat}(\bk)$
with $(t_1,t_2,t_3,t_4,t_5,t_6,b_1,b_2,b_3)=(-1.9,0.9,-1.65,-1.2,0.2842,-1.146,1.9,-0.9,2.029)$.
$\mc{C}_{\rm BS}$ (red) is a circle with a diameter $\dm \in [0,1]$.
$2\theta_{\rm max}$ denotes the apex angle for the solid angle subtended by $\mc{C}_{\rm BS}$. The apex angle is determined by $\dm$: $\cos^2 \theta_{\rm max}=1-\dm^2$.
$\oint d\phi \sqrt{g(\phi)}=2\pi \dm$, which is two times larger than the case of the LBCP.
}
\label{Fig2}
\end{figure*}

{\it \magenta{QBCPs.|}}
Now we consider QBCPs generally described by the Hamiltonian 
\ba
H_{\rm Q}^{(0)}(\bk) = \sum_{a=0}^{3}\sum_{m=0}^{2} v_{a,m}k_x^{m}k_y^{2-m} \sg_a,
\label{eq:generic_quad}
\ea
where $v_{a,m}$ ($a=0,1,2,3$, $m=0,1,2$) are constants. $H_{\rm Q}^{(0)}(\bk)$ has a QBCP
at $\bk=0$ [\fig{Fig2}(a)], and the Berry curvature around it is always zero as $H_{\rm Q}^{(0)}(\bk)$ is a homogeneous-order Hamiltonian.
After successive unitary transformations, $H_{\rm Q}^{(0)}(\bk)$ becomes 
\bg
H_{\rm Q}(\bk) = (b_1 k_x^2 + b_2 k_x k_y + b_3 k_y^2) \sg_0 + t_6 k_y^2 \sg_x \nn \\
+(t_4 k_x k_y  + t_5 k_y^2) \sg_y + (t_1 k_x^2 + t_2 k_x k_y + t_3 k_y^2) \sg_z,
\label{eq:quad_H}
\eg
where $b_{1,2,3}$ and $t_{1,2,\dots,6}$ are real constants~\cite{rhim2019classification}.
Here $\b f(\bk)$ describes a cone in $[f_1(\bk),f_2(\bk),f_3(\bk)]$ space as $\bk$ is varied.
As a result, $\mc{C}_{\rm BS}$ is no longer a circle but a closed loop with an elliptical shape [see \fig{Fig2}(b)].
Contrary to LBCPs, there is no simple expression connecting the quantum metric and $\dm$
for $H_{\rm Q}(\bk)$ in general.
Nevertheless, $\mc{C}_{\rm BS}$ becomes a circle with arbitrary radius when $C_3$ or $C_6$ symmetry exists or a great circle when time-reversal symmetry is further imposed, depending on the symmetry representation.
In such cases, the relevant Hamiltonian can be reduced to the Hamiltonian describing a flat band with a QBCP, by adding a term proportional to the identity matrix with an appropriate coefficient.
As this procedure does not change the wave function and its geometry, the relevant geometric properties are also identical to those of the flat band with a QBCP, which is discussed below.
More details on the QBCPs with rotation and/or time-reversal symmetries are provided in the Supplemental Material~\cite{supple}.

{\it \magenta{Flat band with a QBCP.|}}
Interestingly, the geometric properties of $H_{\rm Q}(\bk)$ can be fully characterized by the quantum metric, if one of the two crossing bands is flat as in \fig{Fig2}(c).
We denote such a flat band Hamiltonian by $H_{\rm flat}(\b k)$.
It is worth noting that $\mc{C}_{\rm BS}$ corresponding to $H_{\rm flat}(\b k)$ is a circle with diameter $\dm$~\cite{supple} as shown in \fig{Fig2}(d).

The quantum metric $g(\phi)$ of $H_{\rm flat}(\b k)$ also has a closed form~\cite{supple}
which is plotted in \fig{Fig2}(e). 
We find $|\mc{C}_{\rm BS}|=\oint d\phi \sqrt{g(\phi)}=2\pi\dm$.
Here the additional multiplication factor 2, compared to \eq{eq:linear_sum}, arises from the fact that $H_{\rm flat}(\b k)=H_{\rm flat}(-\b k)$, thus $\check{\b n}(\bk)$ at $\phi$ and $\phi+\pi$ are identical.
As $\check{\b n}(\bk)$ winds twice for one cyclic change of $\phi$, $\mc{C}_{\rm BS}$ is two-fold degenerate.

A straightforward calculation gives
\ba
\dm = \frac{1}{2\pi} \int d\phi \sqrt{g(\phi)} = \frac{|t_4|}{(2 t_4^2 + 4t_1 t_3 - t_2^2)^{1/2}},
\label{eq:dmax_flat}
\ea
which shows that $\dm>0$ ($\dm=0$) when $t_4\neq 0$ ($t_4=0$)~\cite{rhim2020quantum}. 
The flat band with $\dm>0$ ($\dm=0$) is called a singular (non-singular) flat band~\cite{rhim2019classification}.

The solid angle subtended by $\mc{C}_{\rm BS}$ is $\Omega(\mc{C}_{\rm BS})=4s\pi (1-\cos\theta_{\rm max})$ where the apex angle $2\theta_{\rm max}$ satisfies $\cos\theta_{\rm max}=\sqrt{1-\dm^2}$, and $s={\rm sign}(\frac{t_1 t_6}{t_4})=+1$ ($-1$) indicates the counter-clockwise (clockwise) orientation of $\mc{C}_{\rm BS}$.
Note that the apex angle is defined so that $\theta_{\rm max}$ is not greater than $\pi/2$ as shown in \fig{Fig2}(d).
Hence, the Berry phase is determined by $\dm$ as
\ba
\Phi_B(\mc{C}_{\rm BZ})=2s \pi \sqrt{1-\dm^2}
\label{eq:quad_Berry}
\ea
modulo $2\pi$.
We note that the Berry phase around a QBCP can take any value between $0$ and $2\pi$ (mod $2\pi$).
Although vanishing Berry curvature guarantees path-independent Berry phase, its value depends on the Hamiltonian parameters. 
Also, the quantum metric or the quantum distance is a more useful geometric quantity than the Berry phase for describing BCPs because the Berry phase cannot distinguish singular BCPs with $\dm=1$ and non-singular BCPs with $\dm=0$.
We note that $\dm$ of a QBCP becomes $0$ or $1$, which indicates that the Berry phase is zero modulo $2\pi$, when space-time inversion symmetry exists~\cite{supple}.

\begin{figure}[b!]
\centering
\includegraphics[width=0.45\textwidth]{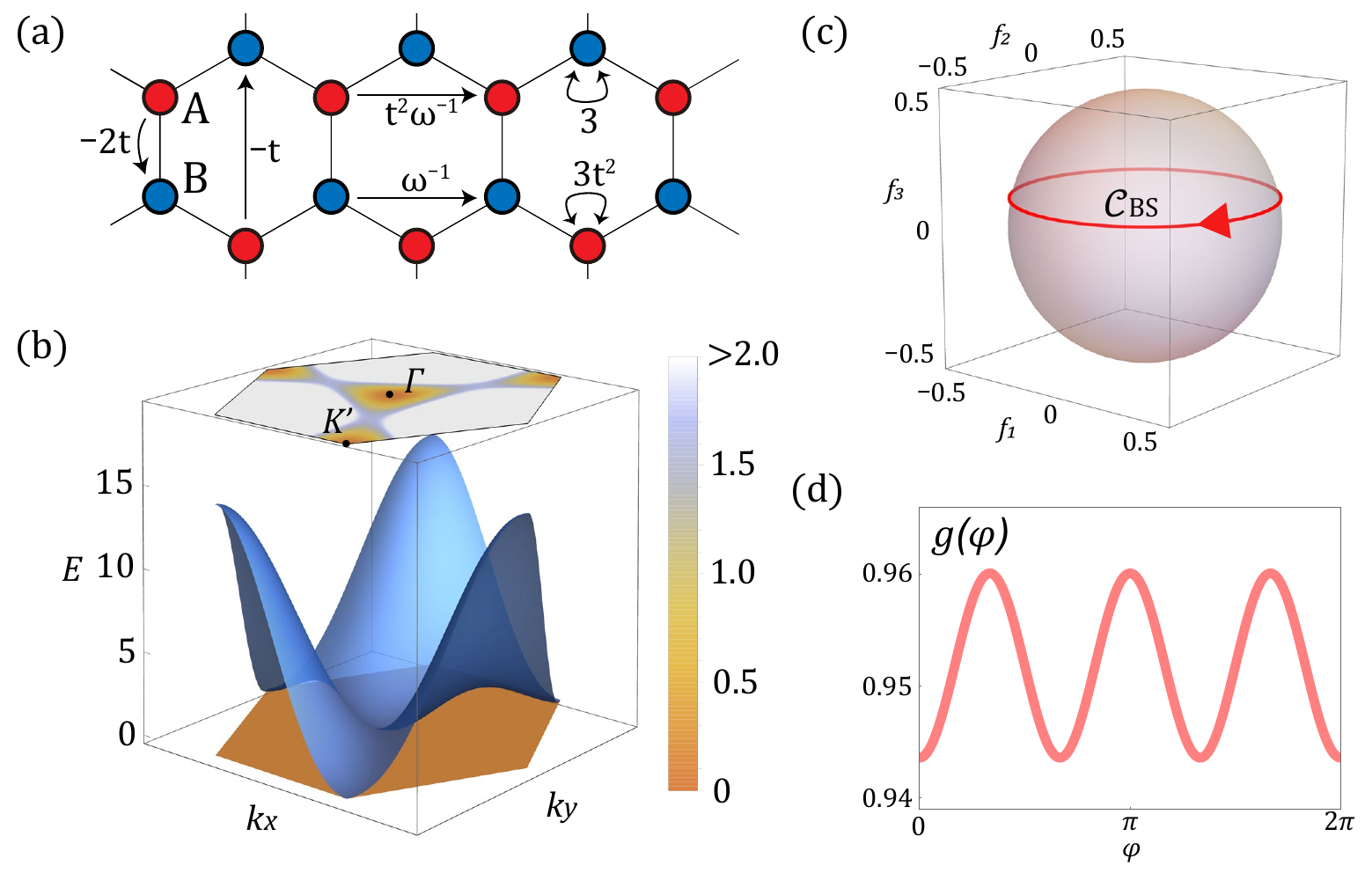}
\caption{
A tight-binding model $H_{\rm double}(\bk)$ exhibiting a flat band with two QBCPs.
(a) The honeycomb lattice with two sublattices $A$ and $B$.
Black arrows denote the hopping interactions.
Other hopping processes related by $C_3$ rotation are not shown for clarity.
(b) The band structure for $t=0.8$ displaying two QBCPs at $\Gamma$ and $K'$ points, respectively.
The energy difference between two bands is indicated by the intensity plot on the top.
(c) The trajectory of the occupied eigenstate on $S^2_{\rm BS}$ as the momentum changes
along a circle enclosing the BCP at $\Gamma$.
A similar circular trajectory with the opposite orientation can be found for the BCP at $K'$.
(d) The quantum metric $g(\phi)$, evaluated along a small circle enclosing
the BCP at $\Gamma$.
}
\label{Fig3}
\end{figure}

{\it \magenta{Tight-binding model.|}}
We construct a lattice model displaying our result on a flat band with QBCP.
This model is defined on the honeycomb lattice including the hoppings up to third nearest-neighbor sites [\fig{Fig3}(a)].
The lattice Hamiltonian with $C_3$ symmetry is given by
\ba
H_{\rm double}(\bk) = \bpm t^2|g(\bk)|^2 & -t\om^* [g(\bk)]^2 \\ -t \om [g^*(\bk)]^2 & |g(\bk)|^2 \epm,
\ea
where $g(\bk) = e^{-\frac{i}{3}(k_1+k_2)}(1+\om e^{i k_1}+\om^{-1} e^{ik_2})$, $(k_1,k_2)=(k_x,\frac{1}{2}k_x+\frac{\sqrt{3}}{2} k_y)$, $\om=e^{\frac{2\pi i}{3}}$, and $t$ is a real parameter. 
%

\begin{figure*}[t!]
\centering
\includegraphics[width=0.75\textwidth]{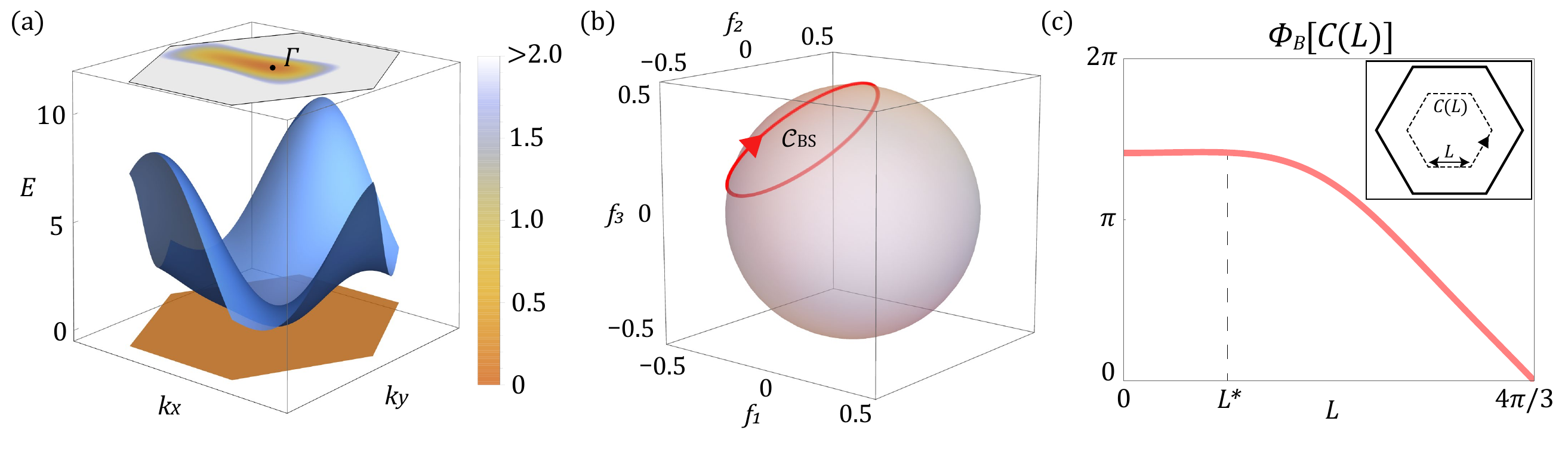}
\caption{
A tight-binding model $H_{\rm single}(\bk)$ exhibiting a flat band with a single QBCP.
(a) The band structure of $H_{\rm single}(\bk)$.
A flat band has a QBCP at $\Gamma$.
The energy difference between two bands is indicated by the intensity plot on the top.
(b) The trajectory of $\ket{\psi_{\rm flat}(\bk)}$ on $S^2_{\rm BS}$ as the momentum changes along a circle enclosing the BCP at $\Gamma$.
(c) The Berry phase along the hexagonal closed loop $C(L)$.
Near the BCP ($L<L^*$), $\Phi_B[C(L)] \simeq \sqrt{2}\pi$ in accordance with \eq{eq:quad_Berry}.
Because of non-zero Berry curvature, $\Phi_B[C(L)]$ changes as $L$ increases.
$C(4\pi/3)$ corresponds to the BZ boundary, and the relevant Berry phase ($\Phi_B[C(4\pi/3)]$) vanishes.
}
\label{Fig4}
\end{figure*}

The relevant band structure exhibits two QBCPs at $\Gamma=(0,0)$ and $K'=(\frac{2\pi}{3},-\frac{2\pi}{\sqrt{3}})$, respectively, [see \fig{Fig3}(b)].
For the QBCP at $\Gamma$ ($K'$), we find $\mc{C}_{\rm BS}$ with $\dm=\frac{2|t|}{t^2+1}$ ($\frac{2|t|}{t^2+1}$) [see \fig{Fig3}(c)], and $\Phi_B= 2\pi \frac{t^2-1}{t^2+1}$ ($-2\pi \frac{t^2-1}{t^2+1}$) satisfying \eq{eq:quad_Berry}.
Hence, the Berry phase can have an arbitrary value depending on $t$.
We note that the model has zero Berry curvature in the BZ except at the BCPs, hence a single BCP at $\Gamma$ with a nonzero Berry phase must accompany another BCP at $K'$ so that the total Berry phase computed along the BZ boundary becomes 0 (mod $2\pi$). 

Meanwhile, a flat band can exhibit a single QBCP with an arbitrary Berry phase when finite Berry curvature exists away from the BCP.
To demonstrate this, we construct a lattice model on the honeycomb lattice.
Explicitly, the Hamiltonian is given by
\ba
H_{\rm single}(\bk) = \frac{1}{4}\bpm |g_2(\bk)|^2 & -g_1(\bk)g_2^*(\bk) \\ -g_1^*(\bk)g_2(\bk) & |g_1(\bk)|^2 \epm,
\label{eq:Htb3}
\ea
where
\ba
g_1(\bk) &= e^{-\frac{i}{3}(k_1+k_2)} \left[2-(1+i)e^{ik_2}-(1-i)e^{-ik_1+ik_2}\right], \nn \\
g_2(\bk) &= e^{-\frac{i}{3}(2k_1-k_2)} \left(2-e^{-ik_2}-e^{ik_1-ik_2}\right).
\ea

The band structure exhibits a single QBCP at $\Gamma=(0,0)$ with $\dm=1/\sqrt{2}$ [see \figs{Fig4}(a) and \ref{Fig4}(b)].
Since the Berry curvature vanishes only near the BCP, the Berry phase $\Phi_B[C(L)]$, computed along the hexagonal closed loop with edges of length $L$, is path-independent when the path is close enough to BCP [see \fig{Fig4}(c)].
In accordance with $\dm=1/\sqrt{2}$, $\Phi_B[C(L)]$ converges to $\sqrt{2}\pi$ as $L \ra 0$.
We note that a single BCP can exist alone as the non-zero Berry curvature makes the Berry phase along the BZ boundary to vanish.

{\it \magenta{Discussion.|}}
We focused on BCP(s) in two-band models.
However, real materials always contain additional bands whose influence can be understood perturbatively.
Namely, starting from the full band structure where two bands form a BCP at $\bk_0$, the effective Hamiltonian $H_{\rm eff}(\b q)$ for the ``crossing bands'' near the BCP at $\bk_0$, where $\bk=\bk_0+\b q$, can be obtained following the L{\"o}wdin perturbation theory~\cite{lowdin1951note}.
Then the geometric properties of $H_{\rm eff}(\b q)$ can be analyzed using our theory as shown in the Supplemental Material~\cite{supple}.

Finally, we note that some lattice models exhibit multi-fold degeneracies where more than two bands cross.
For example, recent studies of three-dimensional and four-dimensional multi-fold fermions have shown that the quantum metric is closely related to the Chern number or tensor monopole charge~\cite{palumbo2018revealing,salerno2020floquet,lin2021dual,chen2020synthetic,lin2021band}.
Understanding the quantum geometry of two-dimensional multi-fold degeneracies, using higher-dimensional Bloch spheres, would be an important problem for future study.

\let\oldaddcontentsline\addcontentsline
\renewcommand{\addcontentsline}[3]{}
\begin{acknowledgments}
Y.H. was supported by the Institute for Basic Science in Korea (Grant No. IBS-R009-D1), Samsung  Science and Technology Foundation under Project Number SSTF-BA2002-06, and the National Research Foundation of Korea (NRF) grant funded by the Korea government (MSIT) (No.2021R1A2C4002773).
J.J. was supported by  
Samsung  Science and Technology Foundation under Project Number SSTF-BA2002-06,
and the National Research Foundation of Korea (NRF) grant funded by the Korea government (MSIT) (No.2021R1A2C4002773).
J.W.R. was supported by Institute for Basic Science in Korea (Grant No. IBS-R009-D1), the National Research Foundation of Korea (NRF) grant funded by the Korea government (MSIT) (Grant No. 2021R1A2C101057211).
B.J.Y. was supported by the Institute for Basic Science in Korea (Grant No. IBS-R009-D1), 
Samsung  Science and Technology Foundation under Project Number SSTF-BA2002-06,
the National Research Foundation of Korea (NRF) grant funded by the Korea government (MSIT) (No.2021R1A2C4002773), and the U.S. Army Research Office and Asian Office of Aerospace Research \& Development (AOARD) under Grant
No. W911NF-18-1-0137.
\end{acknowledgments}

%

\let\addcontentsline\oldaddcontentsline
\clearpage
\onecolumngrid
\begin{center}
\textbf{\large Supplemental Material for ``\ourtitle"}
\end{center}
\setcounter{section}{0}
\setcounter{figure}{0}
\setcounter{equation}{0}
\renewcommand{\thefigure}{S\arabic{figure}}
\renewcommand{\theequation}{S\arabic{equation}}
\renewcommand{\thesection}{S\arabic{section}}
\tableofcontents
\hfill \\
\twocolumngrid

\section{Berry curvature of homogeneous-order Hamiltonian \label{app:Zero_BerryCurv}}
In the main paper, we use the fact that the Berry curvature vanishes except at the band crossing point for a Hamiltonian composed of homogeneous polynomial (a homogeneous-order Hamiltonian for the simplicity)
This can be proved as follows.

Any eigenstate $\ket{\psi(\bk)}$ of a homogeneous-order Hamiltonian in 2D are dependent only on the polar angle $\phi$, thus a derivative on the eigenstates with respect to the radial distance $k=|\bk|$ always gives zero.
Hence, the Berry curvature $F_{k\vph}(\bk)=\der_k A_\phi(\bk) - \der_\phi A_k(\bk)$ must be zero, where $A_i(\bk)=i\bra{\psi(\bk)} \der_i \ket{\psi(\bk)}$ denotes the Berry connection.
Note that $F_{kk}(\bk)=F_{\phi\phi}(\bk)=0$ by definition.

For two-band system, we can use the the following formula for the Berry curvature $F_{ij}(\bk)$:
\ba
F_{ij}(\bk) = -\frac{1}{2|\b f|^3} \b f \cdot \der_i \b f \times \der_j \b f
\label{eq:BerryCurv_f}
\ea
where $\b f=\b f(\bk)$ is defined by \eq{eq:twobandH}.
For a homogeneous-order Hamiltonian,
\ba
k_x \der_x \b f + k_y \der_y \b f - p \b f=0
\ea
holds where $p$ denotes the order of homogeneous polynomials $\b f$.
At $\bk \ne 0$, \eq{eq:BerryCurv_f} leads to
\ba
-2 k_y F_{xy}(\bk) |\b f|^3
=& k_y \ep_{ijl} f_i \der_x f_j \der_y f_l \nn \\
=& \ep_{ijl} f_i \der_x f_j (p f_l - k_x \der_x f_l) \nn \\
=& p (\ep_{ijl} f_i f_l) \der_x f_j - k_x (\ep_{ijl} \der_x f_j \der_x f_l) f_i \nn \\
=& 0
\ea
Thus, $F_{xy}(\bk)=0$ at $\bk\ne0$.

In general, in order to have $F_{xy}(\bk)=0$,
\ba
{\rm Det} (\b f, \der_x \b f, \der_y \b f) = \b f \cdot \der_x \b f \times \der_y \b f
\ea
must vanishes.
We note that 
\ba
\delta_1(\bk) \der_x \b f+ \delta_2(\bk) \der_y \b f + \delta_3(\bk) \b f=0
\label{eq:zeroBerryCond}
\ea 
is a sufficient condition for $F_{xy}(\bk)=0$.
Note that $(\delta_1(\bk),\delta_2(\bk),\delta_3(\bk))$ are given by $(k_x,k_y,-p)$ for a homogeneous-order Hamiltonian of order $p$.

\section{Quantized Berry phase in the presence of symmetry \label{app:symmetry}}
In two-dimensions, space-time inversion $I_{ST}$ symmetry satisfying $I_{ST}^2=1$ is defined either by the combination of time-reversal and inversion symmetries in spinless fermion systems
or by the combination of time-reversal and two-fold rotation about an axis normal to the 2D plane in both spinless and spinful fermion systems.
As $I_{ST}$ is an anti-unitary symmetry which is local in momentum space, it forces the Berry curvature $F_{xy}(\bk)$ to satisfy $F_{xy}(\bk)=-F_{xy}(\bk)$,
which makes $F_{xy}(\bk)=0$ at every momentum $\bk$. Therefore $I_{ST}$ symmetry provides an ideal condition for the path-independent quantized Berry phase.

On the other hand, when the system has a mirror symmetry such as $M_x:(k_x,k_y)\rightarrow(-k_x,k_y)$, a stable linear band crossing point can appear on a mirror invariant line
with $k_x=0$ or $k_x=\pi$. In this case, one conventional way of explaining the quantization of the Berry phase is to perform a line integral along a loop which is $M_x$ symmetric 
so that the Berry curvature integral vanishes due to the relation $F_{xy}(k_x,k_y)=-F_{xy}(-k_x,k_y)$ imposed by $M_x$ symmetry~\cite{fang2015new}.
However, choosing mirror symmetric loops for computing the Berry phase is rather unnatural as the path-independent Berry phase should be independent of the loop shapes.
Moreover, the presence of either space-time inversion or mirror symmetry cannot answer why the quantized value of the Berry phase is exactly equal to $\pi$, although $I_{ST}$ quantizes the Berry phase to be $0$ or $\pi$.

\section{Hamiltonian describing the linear band crossing \label{app:Linear_Crossing_Unitary}}
Here, we derive the Hamiltonian describing the linear band crossing.
We start from the most generic homogeneous-order Hamiltonian of order $p=1$:
\ba
H_{\rm linear}^{(0)}(\bk) = (b_1 k_x + b_2 k_y) \sg_0 + \sum_{a=1}^{3} (v_{ax} k_x + v_{ay} k_y) \sg_a.
\label{eq:generic_linear_app}
\ea
Among six parameters $v_{ai}$ ($a=1,2,3$, $i=x,y$), three can be removed by three successive unitary transformations.
First, we rotate the basis of Hamiltonian by defining $H_{\rm linear}^{(1)}(\bk) = \mc{U}_1 H_{\rm linear}^{(0)}(\bk) \mc{U}_1^{-1}$, where $\mc{U}_1=e^{i \frac{\phi_1}{2} \sg_y}$ and $\phi_1=\tan^{-1}(-v_{3x}/v_{1x})$.
Then, $H_{\rm linear}^{(1)}(\bk)$ can be represented in the same form as \eq{eq:generic_linear_app}, but now with different parameters $v_{ai}$.
Suitable redefinition of the parameters $v_{ai}$ gives $H_{\rm linear}^{(1)}(\bk)$ with $v_{3x}=0$.
Similarly, a unitary transformation with $\mc{U}_2=e^{i \frac{\phi_2}{2} \sg_z}$ and $\phi_2=\tan^{-1}(-v_{1x}/v_{2x})$ defines $H_{\rm linear}^{(2)}(\bk)$ with $v_{1x}=v_{3x}=0$.
Finally, we use $\mc{U}_3=e^{i \frac{\phi_3}{2} \sg_y}$ with $\phi_3=\tan^{-1}(-v_{3y}/v_{1y})$ to remove $v_{3y}$.
Hence, the Hamiltonian with three parameters, $(t_1,t_2,t_3)=(v_{2x},v_{2y},v_{1y})$, are obtained:
\bg
H_{\rm L}(\bk) = (b_1 k_x + b_2 k_y) \sg_0 + t_3 k_y \sg_1 + (t_1 k_x + t_2 k_y) \sg_2.
\label{eq:Hlinear_app}
\eg

\section{Analytic expressions of the quantum metric characterizing the LBCP and QBCP}
In this section, we provide analytic expressions of the quantum metric $g_{\phi\phi}(\phi) \equiv g(\phi)$ for the linear BCP (LBCP) and the quadratic BCP (QBCP).

The LBCP is described by the Hamiltonian $H_L(\bk)$ in \eq{eq:Hlinear_app}.
For $H_L(\bk)$, the quantum metric $g(\phi)$ is given by a closed form,
\ba
g(\phi) = \left( \frac{t_1 t_3/2}{t_1^2 \cos^2 \phi + (t_2^2 + t_3^2) \sin^2 \phi + t_1 t_2\sin{2\phi}} \right)^2,
\ea
where $\phi=\tan^{-1} (k_y/k_x)$.

When a flat band has a band crossing with a quadratic band, the QBPC is described by the Hamiltonian $H_{\rm flat}(\bk)$,
\bg
H_{\rm flat}(\bk) = (b_1 k_x^2 + b_2 k_x k_y + b_3 k_y^2) \sg_0 + t_6 k_y^2 \sg_x \nn \\
+(t_4 k_x k_y  + t_5 k_y^2) \sg_y + (t_1 k_x^2 + t_2 k_x k_y + t_3 k_y^2) \sg_z,
\label{eq:H_quad_app}
\eg
with the condition ${\rm Det} H_{\rm flat}(\bk)=0$ imposing flat dispersion to one of the two crossing band.
For $H_{\rm flat}(\bk)$, the quantum metric $g(\phi)$ is given by a closed form,
\ba
g(\phi)=\left( \frac{t_1 t_4}{2t_1^2 \cos^2 \phi + (2t_1 t_3 + t_4^2) \sin^2 \phi + t_1 t_2 \sin 2\phi} \right)^2.
\ea

\section{Wave function geometry and the Bloch sphere \label{app:Bloch_Sphere}}
The Bloch sphere provides a useful geometric interpretation for two-band system, as introduced in the main text.
In particular, the quantum distance between two eigenstates is given by the straight-line distance between two points on the Bloch sphere corresponding to those eigenstates.
In this section, we discuss the definition and properties of the Bloch sphere.

Let us consider a two-band Hamiltonian
\ba
H(\bk)=f_0(\bk) \sg_0 - \b f (\bk) \cdot \b \sg,
\label{eq:twobandH}
\ea
where $\b f(\bk) = |\b f(\bk)| (\cos \alpha \sin \beta, \sin \alpha \sin \beta, \cos \beta)$ with angle variables $0 \le \alpha <2\pi$ and $0 \le \beta \le \pi$.
The ground state $\ket{\psi_g(\alpha,\beta)}$ is given by
\ba
\ket{\psi_g(\alpha,\beta)} = \bpm \cos \frac{\beta}{2} e^{-i \alpha} \\ \sin \frac{\beta}{2} \epm.
\label{eq:qubit}
\ea
Now, we calculate the quantum distance between $\ket{\psi_g(\alpha,\beta)}$ and $\ket{\psi_g(\alpha+d\alpha,\beta+d\beta)}$, i.e., for the infinitesimal change of $\b \ep = (\alpha,\beta)$,
\ba
d^2(\b \ep, \b \ep+d\b \ep)
=& 1 - |\brk{\psi_g(\b \ep)}{\psi_g(\b \ep+d\b \ep)}|^2 \nn \\
&= \frac{1}{4} d\beta^2 + \frac{\sin^2{\beta}}{4} d\alpha^2 \nn \\
&=\mf{g}_{\ep_I\ep_J}(\b \ep) d\ep_I d\ep_J,
\label{eq:BS_metric}
\ea
where $I,J=1,2$.
The last line of \eq{eq:BS_metric} defines the quantum metric, which is associated with the ground state manifold defined by $\ket{\psi_g(\b \ep)}$,
\bg
\mf{g}_{\alpha \alpha}(\b \ep) = \frac{\sin^2 \beta}{4}, \quad \mf{g}_{\beta \beta}(\b \ep)=\frac{1}{4}, \nn \\
\mf{g}_{\alpha \beta}(\b \ep)=\mf{g}_{\beta \alpha}(\b \ep)=0.
\label{eq:BS_metric_comp}
\eg
The components of the quantum metric in ~\eq{eq:BS_metric_comp} are identical to those for the sphere $S^2$ with a radius $\frac{1}{2}$.
Using this correspondence, we define the Bloch sphere $S^2_{\rm BS}$ with radius $\rbs=\frac{1}{2}$ and a mapping of $\ket{\psi_g(\b \ep)}$ to a point on the Bloch sphere,  $\check{\b n}(\bk) \in S^2_{\rm BS}$,
\ba
\check{\b n}(\bk) = \frac{\b f(\bk)}{2|\b f(\bk)|} = \frac{1}{2} (\cos \alpha \sin \beta, \sin \alpha \sin \beta, \cos \beta).
\label{eq:nvector}
\ea
Thus, the angle variables $\alpha$ and $\beta$ correspond to the polar and azimuthal angles of the Bloch sphere, respectively.
Note that the mapping \eq{eq:nvector} defines the Hopf fibration:
\ba
\check{\b n}(\bk) = \frac{1}{2} \bra{\psi_g(\b \ep)} \b \sg \ket{\psi_g(\b \ep)}
\label{eq:Hopf_fib}.
\ea
We also note that the quantum metric in momentum space $g_{ij}(\bk)$ ($i,j=x,y$) can be obtained by a pullback from the Bloch sphere to momentum space:
\ba
g_{ij}(\bk)
&= \mf{g}_{IJ}(\b \ep) \frac{\der \ep_I}{\der k_i} \frac{\der \ep_J}{\der k_j} \nn \\
&= \der_i \check{\b n}(\bk) \cdot \der_j \check{\b n}(\bk).
\ea

We now consider two quantum states $\ket{\psi_g(\b \ep)}$ and $\ket{\psi_g(\b \ep')}$ with $\b \ep=(\alpha,\beta)$ and $\b \ep'=(\alpha',\beta')$.
These states are mapped to $\check{\b n}(\bk)$ and $\check{\b n}(\bk')$ on the Bloch sphere through \eq{eq:Hopf_fib}.
Then, the quantum distance between $\ket{\psi_g(\b \ep)}$ and $\ket{\psi_g(\b \ep')}$,
\ba
d^2(\b \ep,\b \ep')= 1-|\brk{\psi_g(\b \ep)}{\psi_g(\b \ep')}|^2,
\label{eq:qdis_supp}
\ea
is simply given by the straight-line distance between $\check{\b n}(\bk)$ and $\check{\b n}(\bk')$ on the Bloch sphere, $|\check{\b n}(\bk)-\check{\b n}(\bk')|$.
This can be verified by noticing that
\ba
\ket{\psi_g(\b \ep)} \bra{\psi_g(\b \ep)} = \frac{1}{2} \sg_0 + \check{\b n}(\bk) \cdot \b \sg,
\ea
which can be shown straightforwardly from \eq{eq:qubit}.
Accordingly, \eq{eq:qdis_supp} becomes
\ba
d^2(\b \ep,\b \ep') =& 1-\brk{\psi_g(\b \ep)}{\psi_g(\b \ep')} \brk{\psi_g(\b \ep')}{\psi_g(\b \ep)} \nn \\
=& 1- \bra{\psi_g(\b \ep)} \left[ \frac{1}{2}\sg_0 + \check{\b n}(\bk') \cdot \b \sg \right]\ket{\psi_g(\b \ep)} \nn \\
=& \frac{1}{2} - 2 \check{\b n}(\bk) \cdot \check{\b n}(\bk') \nn \\
=& \left|\check{\b n}(\bk)-\check{\b n}(\bk')\right|^2
\ea
where \eq{eq:Hopf_fib} and $|\check{\b n}(\bk)|^2=|\check{\b n}(\bk)|^2=\frac{1}{4}$ are used in the third and last equalities.

The quantum distance $d(\b \ep,\b \ep')$ in \eq{eq:qdis_supp} is known as the Hilbert-Schmidt distance.
On the other hand, the Fubini-Study distance $|L_{\b \ep,\b \ep'}|$ between $\ket{\psi_g(\b \ep)}$ and $\ket{\psi_g(\b \ep')}$ is given by the integration of the infinitesimal Hilbert-Schmidt distance $d(\b \ep,\b \ep')$ along the geodesic path connecting $\b \ep$ and $\b \ep'$:
\ba
|L_{\b \ep,\b \ep'}|=\int_{\b \ep}^{\b \ep'} \sqrt{\mf{g}_{\ep_I \ep_J}(\b \ep) d\ep_I d\ep_J}.
\ea
Generally, the Hilbert-Schmidt and Fubini-Study distances have different interpretations and these cannot be interchanged.
However, as we study in the main text, the ground state manifold is given by a circle with diameter $\dm$ for wave functions near a linear band crossing, or a quadratic band crossing when one of the crossing band is flat or when the system respects rotation and/or time-reversal symmetries.
In these cases, both notions of distances can be exchanged, and these are simply proportional to each other.

Also, the Berry phase is expressed as an geometric quantity of the Bloch sphere.
For a closed loop $\mc{C}_{\rm BZ}$ in momentum space, the Berry phase (modulo $2\pi$) is given by
\ba
\Phi_B(\mc{C}_{\rm BZ})
&= \oint_{\mc{C}_{\rm BZ}}i \bra{\psi_g(\alpha,\beta)} \nabla_\bk \ket{\psi_g(\alpha,\beta)} \cdot d\bk \nn \\
&= \frac{1}{2} \oint_{\mc{C}_{\rm BZ}} \left(1+\cos\beta\right) \, \nabla_\bk \alpha \cdot d\bk \nn \\
&= -\frac{1}{2} \oint_{\mc{C}_{\rm BZ}} \left(1-\cos\beta\right) \, \nabla_\bk \alpha \cdot d\bk \nn \\
&= -\frac{1}{2} \Omega(\mc{C}_{\rm BS}).
\label{eq:BP_Solid}
\ea
Here, $\Omega(\mc{C}_{\rm BS})$ is the solid angle subtended by $\mc{C}_{\rm BS}$, and a closed loop $\mc{C}_{\rm BS}$ is a mapping of $\mc{C}_{\rm BZ}$ onto the Bloch sphere.

We comment that a quantum state can be mapped on the Bloch sphere by using the Hopf fibration in \eq{eq:Hopf_fib} without mentioning the Hamiltonian.
Also, the quantum state is not necessary to be the ground state.

In the main text, we study the geometric quantities of band crossing points, when $\mc{C}_{\rm BS}$ on the Bloch sphere, a trajectory determined by $\check{\b n}(\bk)$, is a circle.
In this case, the quantum metric, the maximum quantum distance, and the Berry phase are related by the closed-form expressions.

First, we note that the diameter of $\mc{C}_{\rm BS}$ is equal to the maximal quantum distance $\dm$, since the quantum distance of two eigenstates is equivalent to the straight line distance between two corresponding points on $\mc{C}_{\rm BS}$.
Meanwhile, the solid angle subtended by $\mc{C}_{\rm BS}$ has its apex angle $2\theta_{\rm max} \le \pi$, which is determined by $\dm$: $\cos^2 \theta_{\rm max} = 1-(\frac{\dm}{2r_{\rm BS}})^2=1-\dm^2$.
In addition, we must take into account a winding number $N_w$ of $\mc{C}_{\rm BS}$.
The sign of $N_w$ is an orientation of $\mc{C}_{\rm BS}$ and the absolute value of $N_w$ indicates how many times $\mc{C}_{\rm BS}$ is overlapped.

Hence, the length of $\mc{C}_{\rm BS}$ is expressed as $|\mc{C}_{\rm BS}|=\pi |N_w| \dm$.
Equivalently, $|\mc{C}_{\rm BS}|$ is also given by the Fubini-Study distance, which is given by the integration of quantum metric, $\oint d\phi \, \sqrt{g_{\phi\phi}(\phi)}$.
Here, $g_{\phi\phi}(\phi)$ denotes the angular part of quantum metric and $\phi=\tan^{-1}(k_y/k_x)$.
These two different expressions for $|\mc{C}_{\rm BS}|$ relate the maximum quantum distance and the angular integral of quantum metric:
\ba
\dm = \frac{1}{\pi |N_w|} \oint d\phi \, \sqrt{g_{\phi \phi}(\phi)}.
\label{eq:dm_metric}
\ea

Now, we show that the Berry phase $\Phi_B(\mc{C}_{\rm BZ})$ is determined by $\dm$, because the apex angle of solid angle $\Omega(\mc{C}_{\rm BS})$ depends on $\dm$.
Hence $\Omega(\mc{C}_{\rm BS})=2\pi N_w (1-\cos \theta_{\rm max})$.
According to \eq{eq:BP_Solid}, the Berry phase is expressed as
\ba
\Phi_B(\mc{C}_{\rm BZ}) =  -\pi N_w (1-\sqrt{1-\dm^2}).
\label{eq:dm_Berry}
\ea
Two equations, Eqs.~\eqref{eq:dm_metric} and \eqref{eq:dm_Berry}, reproduce the results on LBCP and QBPC discussed in the main text.

\section{Wave function geometry of quadratic band crossing \label{app:Cone_Eq}}
In this section, we discuss the wave function geometry of the QBCP.
As discussed in Appendix~\ref{app:Bloch_Sphere}, the wave function geometry can be analyzed visually by studying $\mc{C}_{\rm BS}$ on the Bloch sphere.
Let us fist consider the Hamiltonian describing the quadratic band crossing point:
\bg
H_{\rm Q}(\bk) = (b_1 k_x^2 + b_2 k_x k_y + b_3 k_y^2) \sg_0 - \b f(\bk) \cdot \b \sg,
\label{eq:quadH_supp}
\eg
where $\b f (\bk)=-(t_6 k_y^2, t_4 k_x k_y+t_5 k_y^2, t_1 k_x^2 + t_2 k_x k_y + t_3 k_y^2)$.
First, let us assume that $t_1 t_4 t_6 \ne 0$.
Thus, $k_{x,y}$ can be expressed by $\b f(\bk)$ as follows,
\bg
k_x^2 = -\frac{1}{t_1}\left(f_3-\frac{t_2}{t_4}f_2+\left(\frac{t_2 t_5}{t_4 t_6}-\frac{t_3}{t_6}\right) f_1 \right), \nn \\
k_y^2 = -\frac{1}{t_6}f_1, \quad
k_x k_y =-\frac{1}{t_4}(f_2-\frac{t_5}{t_{6}}f_1),
\eg
where $f_{1,2,3}=f_{1,2,3}(\bk)$ for simplicity.
From $k_x^2 k_y^2 = (k_x k_y)^2$, we obtain an equation describing the surface formed by $f_{1,2,3}$, $f_i \mc{M}_{ij} f_j=0$ $(i,j=1,2,3)$, where $\mc{M}$ is given by 
\bg
\bpm t_1 t_5^2 + t_3 t_4^2 - t_2 t_4 t_5 & -t_1 t_5 t_6 + \frac{1}{2} t_2 t_4 t_6 & -\frac{1}{2} t_4^2 t_6 \\
-t_1 t_5 t_6 + \frac{1}{2} t_2 t_4 t_6 & t_1 t_6^2 & 0 \\
-\frac{1}{2} t_4^2 t_6 & 0 & 0 \epm.
\label{eq:M_eq}
\eg
$\mc{M}$ is a real symmetric matrix and diagonalizable by an orthogonal matrix $O$:
\ba
\mc{M}=O^T \mc{D} O \quad {\rm and} \quad \mc{D}={\rm Diag}(\xi_1,\xi_2,\xi_3).
\ea
We obtain an equation satisfied by $\b f(\bk)$,
\ba
\xi_1 \tilde{f}_1^2+\xi_2 \tilde{f}_2^2+\xi_3 \tilde{f}_3^2=0,
\label{eq:f_eq}
\ea
where $\tilde{f}_i=O_{ij} f_j$.
Since ${\rm Det} \mc{M}=-\frac{1}{4} t_1 t_4^4 t_6^4 \ne 0$, all $\xi_i$ are non-zero.
We note that one of $\xi_i$ must have a different sign than the other two, because otherwise \eq{eq:f_eq} implies $\tilde{\b f}(\bk)=0$.
Hence, \eq{eq:f_eq} describes the elliptic cone.
The shape of $\mc{C}_{\rm BS}$ can be obtained by a projection of the elliptic cone onto the Bloch sphere ($\tilde{f}_1^2+\tilde{f}_2^2+\tilde{f}_3^2=\frac{1}{4}$).
We call such shape an ``elliptical shape" in this paper.
When $t_1 t_4 t_6=0$, it can be shown that $\mc{C}_{\rm BS}$ is an arc with arbitrary central angle.

\subsection{Shape of \texorpdfstring{$\mc{C}_{\rm BS}$}{CBS} in the presence of \texorpdfstring{$C_n$}{Cn} rotation and/or time-reversal symmetry}
In the absence of spin-orbit coupling, $C_n$-rotation operator satisfies $(C_n)^n=1$.
For two-band system, $C_n$ operator is represented by
\ba
C_n = \bpm e^{i\frac{2\pi a}{n}} & 0 \\ 0 & e^{i\frac{2\pi b}{n}} \epm \quad (a,b=0,\dots,n-1).
\ea
We note that the representations with $(a,b)$ and $(a',b')$ are equivalent up to unitary transformation or $U(1)$ phase if $a-b=\pm(a'-b')$ (mod $n$).

A generic Hamiltonian describing two-band system can be expressed by
\ba
H(\bk) = \sum_{a=0,3,\pm} h_a(\bk) \sg_a,
\ea
where $k_\pm = k_+ \pm i k_y$ and $\sg_\pm = \frac{1}{2} (\sg_x \pm i \sg_y)$.
Note that $h_{0,3}(\bk)$ are real functions of $\bk$ while $h_\pm(\bk)=h_\mp^*(\bk)$ is complex function of $\bk$.
Also, $h_a(\bk)$ can be expanded in a power series with respect to $k_{\pm}$:
\bg
h_0(\bk) = \sum_{i,j\ge0} A_{ij} k_+^i k_-^j, \quad
h_3(\bk) = \sum_{i,j\ge0} B_{ij} k_+^i k_-^j, \nn \\
h_+(\bk) = \sum_{i,j\ge0} C_{ij} k_+^i k_-^j,
\label{eq:kp_coeff}
\eg
where $A_{ij}, B_{ij} \in \mathbb{R}$ and $C_{ij} \in \mathbb{C}$.
Then, a symmetry relation, 
$C_n H(\bk) C_n^{-1} = H(C_n\bk) = H(e^{i\frac{2\pi}{n}}k_+,e^{-i\frac{2\pi}{n}}k_-)$,
imposes the following conditions:
\bg
h_{0,3}(\bk) = h_{0,3}(e^{i\frac{2\pi}{n}}k_+,e^{-i\frac{2\pi}{n}}k_-), \nn \\
h_+(\bk) = e^{-i \frac{2\pi(a-b)}{n}} h_+(e^{i\frac{2\pi}{n}}k_+,e^{-i\frac{2\pi}{n}}k_-).
\eg
Hence, $A_{ij}=B_{ij}=0$ if  $i-j \notin n \mathbb{Z}$, and $
C_{ij}=0$ if $i-j-a+b \notin n \mathbb{Z}$.
Up to the second order of $k$, $h_{0,3}(\bk)$ can be expanded as,
\ba
h_0(\bk) = A_{00}+A_{11} k^2, \quad
h_3(\bk) = B_{00}+B_{11} k^2,
\ea
regardless of $a$ and $b$.

Here, we work out $C_6$ rotation with $a-b=2$ explicitly as an example.
Since $C_{ij}$ vanishes unless $i-j-a+b \in 6 \mathbb{Z}$, thus we obtain $h_+(\bk) = C_{20} k_+^2 + O(k^3)$.
Hence, a quadratic Hamiltonian, \textit{i.e.} a homogeneous-order Hamiltonian of order 2, is expressed as
\ba
H(\bk)=
k^2
\bpm
A_{11}+B_{11} & C_{20} e^{2i\phi}\\
C_{20}^* e^{-2i\phi} & A_{11}-B_{11}
\epm,
\label{eq:kp_C6}
\ea
where $k=|\bk|$ and $\phi=\tan^{-1}(k_y/k_x)$.
In this case, $\mc{C}_{\rm BS}$, formed by $\check{\b n}(\bk)$ on the Bloch sphere, is a circle with radius $\frac{|C_{20}|}{2 (B_{11}^2+|C_{20}|^2)^{1/2}}$.

Straightforwardly, one can find the conditions that $C_n$-symmetric and quadratic Hamiltonian exhibits $\mc{C}_{\rm BS}$ as a circle as follows:
\ba
C_3: \, a-b=1 \quad {\rm and} \quad C_6: \, a-b=2.
\label{eq:Cn_cond1}
\ea
In other cases, $\mc{C}_{\rm BS}$ is a point or an elliptical shape depending on the representations.

Now, we further impose time-reversal symmetry $T$.
In the absence of spin-orbit coupling, $T$ operator satisfies $T^2=1$ and $[C_n,T]=0$.
Let us consider a case when $T=\mc{K}$.
Then, only $(a,b)=(0,0)$ is allowed for $C_{2m+1}$ while $(a,b)=(0,0)$ and $(0,m)$ are allowed for $C_{2m}$.
In all these cases except $C_2$ rotation with $(a,b)=(0,0)$ and $C_4$ rotation with $(a,b)=(0,2)$, $\mc{C}_{\rm BS}$ is a point.
On the other hand, $\mc{C}_{\rm BS}$ appears as an arc when $(a,b)=(0,0)$ for $C_2$ and $(a,b)=(0,2)$ for $C_4$.
We also note that in the former case, $\mc{C}_{\rm BS}$ may be great circle depending on the parameters.

On the other hand, when $T=\sg_x\mc{K}$, $C_n$ rotation with $(a,b)=(a,-a)$ is compatible with time-reversal symmetry.
Note that this representation is equivalent to 
\bg
C_n=
\bpm \cos \frac{2\pi a}{n} & -\sin \frac{2\pi a}{n} \\
\sin \frac{2\pi a}{n} & \cos \frac{2\pi a}{n} \epm, \quad
T=\mc{K}
\eg
up to unitary transformation.
Time-reversal symmetry $T=\sg_x \mc{K}$ imposes $h_3(\bk)=-h_3(-\bk)$ and $h_{0,+}(\bk)=h_{0,+}(-\bk)$.
Hence, $h_3(\bk)$ vanishes for the quadratic Hamiltonians.
Combining the previous result for $C_n$ rotation, we find that $\mc{C}_{\rm BS}$ is given by great circle and exhibits the maximal quantum distance $\dm=1$ in the following cases:
\ba
C_{3,4}: \, a=1 \quad {\rm and} \quad C_6: \, a=1,2.
\label{eq:Cn_cond2}
\ea
In the case of $C_4$ rotation with $a=1$, an additional condition that $C_{20} \ne \pm C_{02}$ in \eq{eq:kp_coeff} is required.
Note that the relevant quadratic Hamiltonian is expressed as
\ba
H(\bk)=
k^2
\bpm
A_{11} & C_{20} e^{2i\phi}+C_{02} e^{-2i\phi} \\
c.c. & A_{11}
\epm,
\label{eq:kp_C4}
\ea
where $c.c.$ denotes the complex conjugation.

In summary, when the representation of $C_n$ rotation is given by \eq{eq:Cn_cond1} or \eqref{eq:Cn_cond2}, $\mc{C}_{\rm BS}$ is a circle so that the maximal quantum distance $\dm$ fully characterizes the wave function geometry.
This result is revisited in the next subsection.

\subsection{Shape of \texorpdfstring{$\mc{C}_{\rm BS}$}{CBS} of a flat band with a QBCP}
One of the quadratic bands of $H_{\rm Q}(\bk)$ becomes completely flat if ${\rm Det} H_{\rm Q}(\bk)=0$.
This conditions is equivalent to
\ba
2t_1t_5=t_2t_4, \quad t_2^2 t_6^2=t_5^2(t_4^2-t_2^2+4t_1t_3),
\label{eq:flatCond}
\ea
thus only four parameters (say, $t_{1,2,3,4}$) are independent.
Now, we denote the resulting Hamiltonian by $H_{\rm flat}(\bk)$.
As discussed in the main text, for a flat band wave function of $H_{\rm flat}(\bk)$, $\mc{C}_{\rm BS}$ becomes a circle with a diameter $\dm$,
\ba
\dm = \frac{|t_4|}{(2 t_4^2 + 4t_1 t_3 - t_2^2)^{1/2}}.
\ea
Here, we derive this result by obtaining an equation satisfied by $\b f(\bk)$.
Let us recall the equation $f_i \mc{M}_{ij} f_j=0$ where $\mc{M}$ is given in \eq{eq:M_eq}.
If we impose the flat band condition on \eq{eq:flatCond}, we have $f_i \mc{M}'_{ij} f_j=0$ where
\ba
\mc{M}'= \bpm 1 - \xi^2 & 0 & -\xi \\
0 & 1 & 0 \\
-\xi & 0& 0 \epm
\quad {\rm and} \quad
\xi=\frac{t_4^2}{2t_1 t_6}.
\ea
The eigenvalues of $\mc{M}'$ are evaluated as 1, 1, and $-\xi^2$.
This implies that the surface formed by $f_{1,2,3}$ is a circular cone.
Explicitly, the cone in $(f_1,f_2,f_3)$ space satisfies $\tilde{f}_1^2+\tilde{f}_2^2-\xi^2\tilde{f}_3^2=0$ where $(\tilde{f}_1,\tilde{f}_2,\tilde{f}_3)=(-f_1\cos\theta+f_3\sin\theta,f_2,f_1\sin\theta+f_3\cos\theta)$ and $\theta=\tan^{-1}\xi$.
Hence, the intersection between the cone ($\tilde{f}_1^2+\tilde{f}_2^2-\xi^2\tilde{f}_3^2=0$) and the Bloch sphere ($\tilde{f}_1^2+\tilde{f}_2^2+\tilde{f}_3^2=\frac{1}{4}$) determines the shape of $\mc{C}_{\rm BS}$:
\ba
\tilde{f}_1^2+\tilde{f}_2^2 = \frac{\xi^2}{4(1+\xi^2)}
\quad {\rm and} \quad
\tilde{f}_3^2 = \frac{1}{4(1+\xi^2)},
\ea
which describes a circle with diameter $\dm$,
\ba
\dm=2 \sqrt{\tilde{f}_1^2+\tilde{f}_2^2}=\frac{|t_4|}{(2t_4^2+4t_1 t_3 -t_2^2)^{1/2}}.
\ea

In the previous subsection, we obtain the conditions Eqs.~\eqref{eq:Cn_cond1} and \eqref{eq:Cn_cond2} for a quadratic Hamiltonian to have $\mc{C}_{\rm BS}$ as a circle, in the presence of $C_n$ and/or $T$.
This result can also be understood from the fact that $\mc{C}_{\rm BS}$ is always a circle for a flat band with a QBCP.
When a symmetry representation of $C_n$ and $T$ meets the conditions \eq{eq:Cn_cond1} or \eqref{eq:Cn_cond2}, the quadratic Hamiltonian $H(\bk)$ can be made to satisfy ${\rm Det} H(\bk)=0$ by adding a term proportional to identity matrix $\epsilon_0(\bk) \sg_0$ to $H(\bk)$.
[We note one exception: a $C_4$ rotation with $a=1$ in \eq{eq:Cn_cond2} where the corresponding quadratic Hamiltonian is given by \eq{eq:kp_C4}.]
Because the term $\epsilon_0(\bk) \sg_0$ does not change neither a wave function of occupied state nor $f(\bk)$ in \eq{eq:quadH_supp}, $\mc{C}_{\rm BS}$ is given as a circle like as the case of flat band.

\subsection{\texorpdfstring{$\dm$}{dmax} in the presence of space-time-inversion symmetry}
Space-time-inversion $I_{ST}$ is a combined operation of space inversion and time reversal.
In the absence of spin-orbit coupling, $I_{ST}$ satisfies $I_{ST}^2=1$.
Here, in order to exploit the previous results, we choose $I_{ST}=\sg_z \mc{K}$ where $\mc{K}$ denotes the complex conjugation.
Thus, $I_{ST}$ imposes that
\ba
I_{ST} H(\bk) I_{ST}^{-1} = \sg_z H^*(\bk) \sg_z = H(\bk).
\label{eq:PTcondition}
\ea

\eq{eq:PTcondition} implies that $f_1(\bk)=0$, thus $\check{\b n}_1(\bk)=0$.
Accordingly, $\mc{C}_{\rm BS}$ formed by $\check{\b n}(\bk)$ lies in the $(f_2,f_3)$-plane.
For generic $f_2(\bk)$ and $f_3(\bk)$, $\mc{C}_{\rm BS}$ will be given by an arc with an arbitrary central angle.

Now, we show that $\dm$ is quantized to $0$ or $1$ in the flat band condition.
A general form of $I_{ST}$-symmetric Hamiltonian describing a QBCP is given by
\bg
H_{\rm quad}^{(0)}(\bk) = (b_1 k_x^2 + b_2 k_x k_y + b_3 k_y^2) \sg_0 \nn \\
+ \sum_{a=y,z} (v_{a1} k_x^2 + v_{a2} k_x k_y +v_{a3} k_y^2) \sg_a.
\eg
We can set $v_{21}=0$ using a unitary transformation, $H_{\rm quad}(\bk) = \mc{U} H_{\rm quad}^{(0)}(\bk) \mc{U}^{-1}$, where $\mc{U}=e^{i \frac{\theta}{2} \sg_x}$ and $\theta=\tan^{-1}(-v_{21}/v_{31})$.
Redefining the parameters, $H_{\rm quad}(\bk)$ can be expressed as
\bg
H_{\rm quad}(\bk)=(b_1 k_x^2 + b_2 k_x k_y + b_3 k_y^2) \sg_0 \nn \\
+ (t_4 k_x k_y + t_5 k_y^2) \sg_y + (t_1 k_x^2 + t_2 k_x k_y + t_3 k_y^2) \sg_z.
\eg
The flat band condition ${\rm Det} H_{\rm quad}(\bk)=0$ imposes that $t_4=0$ or $t_4^2=t_2^2-4t_1 t_3$. 
Hence, $\dm=|t_4|(2 t_4^2 + 4t_1 t_3 - t_2^2)^{-1/2}$ is given by $0$ and $1$, for each case.

In the presence of $I_{ST}$ symmetry, we note that the Berry phase is quantized to be 0 modulo $2\pi$ for the QBCP regardless of the flat band condition.

\section{Tight-binding models \label{app:TB_Model}}
In this section, we discuss tight-binding models describing a flat band with quadratic band crossings.
The tight-binding Hamiltonian is defined by
\ba
\hat H_{\rm lattice}
\coloneqq& \sum_{\b R} \sum_{i,j} \sum_{\mu,\nu \in \mathbb{Z}} t_{\mu\nu}^{ij} c^\dg_{i,\b R+\mu\b a_1 + \nu\b a_2} c_{j,\b R},
\ea
where $\b R$ is a unit cell position, $\b a_{1,2}$ are the lattice vectors, and $c_{i,\b R}$ ($c_{i,\b R}^\dagger$) denotes the annihilation (creation) operators for electrons at the $i$th sublattice site in unit cell $\b R$, respectively.
Each element in the hopping matrix $t_{\mu\nu}^{ij}$ describes the strength of hopping interaction from $[j,\b R]$ (sublattice site $j$ in unit cell located at $\b R$) to $[i,\b R+\mu\b a_1+\nu\b a_2]$.

Let us consider the Fourier transformation,
\ba
c_{i,\b R} = \frac{1}{\sqrt{N_{\rm cell}}} \, \sum_{\b k} e^{i \bk \cdot (\b R+\b x_i)} c_{i,\bk},
\ea
where $N_{\rm cell}$ is the number of unit cells and $\b x_i$ denotes the position of sublattice site $i$.
Then, the tight-binding Hamiltonian in momentum space $H(\bk)$ can be obtained by
\bg
\hat H_{\rm lattice} = \sum_{\b k} \sum_{i,j} c^\dg_{i,\b k} H(\bk)_{ij} c_{j,\b k}, \nn \\
H(\bk)_{ij} = \sum_{\mu,\nu \in \mathbb{Z}} t_{\mu\nu}^{ij} e^{-i \bk \cdot (\mu \b a_1 + \nu \b a_2 + \bx_i - \bx_j)}.
\label{eq:latticeH_Fourier}
\eg
From \eqref{eq:latticeH_Fourier}, $t_{\mu\nu}^{ij}$ can be read off from $H(\bk)$:
\ba
t_{\mu\nu}^{ij} = \frac{1}{N_{\rm cell}} \sum_{\bk} e^{i \bk \cdot (\mu \b a_1 + \nu \b a_2 + \bx_i - \bx_j)} H(\bk)_{ij}.
\label{eq:hoppingmatrix}
\ea

\subsection{Models exhibiting a flat band with two QBCPs}
Now, we recall the tight-binding model introduced in the main text.
The model is defined on the honeycomb lattice with two sublattice sites located at $\bx_1=(\frac{1}{2},\frac{1}{2\sqrt{3}})$ and $\bx_2=(\frac{1}{2},-\frac{1}{2\sqrt{3}})$.
The lattice vectors are $\b a_1=(1,0)$ and $\b a_2=(\frac{1}{2},\frac{\sqrt{3}}{2})$.

The tight-binding Hamiltonian $H_{\rm double}(\bk)$ is given by
\ba
H_{\rm double}(\bk) = \bpm t^2|g(\bk)|^2 & -t\om^* [g(\bk)]^2 \\ -t \om [g^*(\bk)]^2 & |g(\bk)|^2 \epm,
\label{eq:Htb1}
\ea
where $g(\bk) = e^{-\frac{i}{3}(k_1+k_2)}(1+\om e^{i k_1}+\om^{-1} e^{ik_2})$, $(k_1,k_2)=(k_x,\frac{1}{2}k_x+\frac{\sqrt{3}}{2} k_y)$, and $\om=e^{\frac{2\pi i}{3}}$.
This model is symmetric under $C_3$ rotation:
\ba
U(C_3) H_{\rm double}(\bk) U(C_3)^{-1} = H_{\rm double}(C_3 \bk)
\ea
where $U(C_3)={\rm Diag}(\om,\om^{-1})$.

An unnormalized wave function of flat band is given by
\ba
\ket{\psi_{\rm flat}(\bk)} = (g(\bk), t\om g^*(\bk))^T.
\ea
The compact localized state (CLS)~\cite{sutherland1986localization,bergman2008band,maimaiti2017compact,read2017compactly} corresponding to $\ket{\psi_{\rm flat}(\bk)}$ can be obtained by the Fourier transformation,
\ba
\ket{w(\b R)} = \sum_{\b R',i} A_i(\b R-\b R') c^\dg_{i,\b R'}
\ea
where $A_i(\b R-\b R') =N_{\rm cell}^{-1} \sum_\bk e^{-i\bk \cdot (\b R-\b R'-\bx_i)} \ket{\psi_{\rm flat}(\bk)}_i$.
Explicitly, the CLS $\ket{w(\b R)}$ is expressed as
\bg
\ket{w(\b R)} = c^\dg_{1,\b R}+\om c^\dg_{1,\b R-\b a_1}+\om^{-1} c^\dg_{1,\b R-\b a_2} \nn \\
+t c^\dg_{2,\b R}+ t\om c^\dg_{2,\b R-\b a_1}+ t\om^{-1} c^\dg_{2,\b R-\b a_1+\b a_2}.
\eg

The band structure exhibits two QBCPs at $\Gamma=(0,0)$ and $K'=(\frac{2\pi}{3},-\frac{2\pi}{\sqrt{3}})$ as shown in Fig.~\magenta{3} in the main text.
This can seen from the fact that wave function of flat band $\ket{\psi_{\rm flat}(\bk)}$ vanishes at each $\Gamma$ and $K'$.
According to Refs.~\onlinecite{bergman2008band,rhim2019classification}, this implies a band crossing at there.

In this model, the Berry curvature $F_{xy}(\bk)$ vanishes in the whole Brillouin zone (BZ).
Hence, the Berry phase corresponding to each BCP is path independent.
The geometric quantities such as the Berry phase $\Phi_B$ and the maximal quantum distance $\dm$ can be obtained analytically using the continuum Hamiltonian near each BCP.
We find that $(\dm,\Phi_B)=(\frac{2|t|}{t^2+1},\frac{t^2-1}{t^2+1})$ for the BCP at $\Gamma$ and $(\frac{2|t|}{t^2+1},-\frac{t^2-1}{t^2+1})$ for the BCP at $K'$.

Now, we introduce another tight-binding model in which $\dm$ can be modulated by the parameters in the model.
We briefly apply a lattice regularization, $k_i \rightarrow \sin k_i$, to \eq{eq:quadH_supp}, for a direct comparison with the continuum Hamiltonian in \eq{eq:quadH_supp}.
Thus, we obtain a tight-binding model, $H(\bk) = h_0(\bk) \sg_0 + \b h(\bk) \cdot \b \sg$, where $h_{0,1,2,3}(\bk)$ are given by
\ba
h_0(\bk)=&b_1\sin^2 k_x + b_2\sin k_x \sin k_y + b_3\sin^2 k_y, \nn \\
h_1(\bk)=&t_6 \sin^2 k_y, \nn \\
h_2(\bk)=&t_4 \sin k_x \sin k_y + t_5 \sin^2 k_y, \nn \\
h_3(\bk)=&t_1\sin^2 k_x + t_2\sin k_x \sin k_y + t_3\sin^2 k_y.
\label{eq:latticeH_components1}
\ea
Note that $h_{1,2,3}(\bk)$ satisfy the sufficient condition for having zero Berry curvature, \eq{eq:zeroBerryCond}, where $(\delta_1(\bk),\delta_2(\bk),\delta_3(\bk))=(\tan{k_x},\tan{k_y},-2)$.
The Hamiltonian exhibits quadratic band crossings at $(0,0)$, $(\pi,0)$, $(0,\pi)$, and $(\pi,\pi)$.
Note that Eqs.~\eqref{eq:hoppingmatrix} and ~\eqref{eq:latticeH_components1} imply that only the hoppings between the next-nearest neighbors are non-zero.
Hence, the lattice model can be divided into two subsystems that do not interact with each other.
The tight-binding Hamiltonian for one subsystem can be obtained by a redefinition of $k_{x,y}$: $k_x \rightarrow \frac{k_x+k_y}{2} $ and $k_y \rightarrow \frac{k_x-k_y}{2}$.
Then, the BCPs occur only at $(0,0)$ and $(\pi,\pi)$.

Now, let us set the parameters to $(t_1,t_3,t_4,t_6,b_1,b_3)=(2,\delta,2,\sqrt{1+2\delta},2,1+\delta)$ and zero for the other parameters.
Note that one of two crossing bands is completely flat, ${\rm Det} H(\bk)=0$, with these values of parameters.
Then, $\dm$ is equal to $(2+2\delta)^{-1/2}$ at each BCP, and the corresponding Berry phases are given by $\pm \pi (\frac{2+4\delta}{1+\delta})^{1/2}$.
In this way, the maximal quantum distance and the Berry phase characterizing the BCPs can be modulated by $\delta$.
For simplicity, we set $\delta=0$ from now on.
Accordingly, \eq{eq:latticeH_components1} becomes
\ba
&h_1(\bk)=\frac{1}{2} - \frac{1}{2} \cos(k_x-k_y), \nn \\
&h_2(\bk)=-\cos k_x+\cos k_y, \nn \\
&h_3(\bk)=1-\cos(k_x+k_y).
\label{eq:Htb2}
\ea
At both BCPs, the corresponding maximal quantum distances are given by $\dm=\frac{1}{\sqrt{2}}$.
While, the Berry phase of BCP at the $(0,0)$ is $-\sqrt{2}\pi$ and one at the $(\pi,\pi)$ is $\sqrt{2}\pi$.

In the both models in Eqs.~\eqref{eq:Htb1} and \eqref{eq:Htb2}, the Berry phase around the boundary of BZ is zero, since the Berry phases of BCPs cancel each other out.
More generally, we conclude that the QBCP with non-zero Berry phase (modulo $2\pi$) cannot exist alone in the lattice model when the Berry curvature $F_{xy}(\bk)$ vanishes in the whole BZ.
Such BCP must accompanies other BCP(s).
It can be shown as follows.

Let us suppose that there are BCPs at $\bk^*_i$ in the BZ.
Then, the Berry phase characterizing the $i$th BCP is given by an integration of Berry connection along the loop $C_i$ enclosing the crossing point $\bk^*_i$: $\Phi_B(\bk^*_i)=\oint_{C_i} d\bk \cdot \b A(\bk)$.
Using the Stokes theorem, the total Berry phase of BCPs can be expressed as
\ba
\sum_i \Phi_B(\bk^*_i)
&=-\int_{BZ-\cup D_i} d^2k \, \hat{z} \cdot \b \der \times \b A(\bk) \nn \\
&= -\int_{BZ-\cup D_i} d^2k \, F_{xy}(\bk),
\ea
where $D_i$ denotes a disk bounded by $C_i,$ and we use the fact that the Berry phase along BZ boundary is zero.
Since $F_{xy}(\bk)=0$, the total Berry phase of BCPs must be zero.
This implies that a BCP with non-zero Berry phase cannot exist alone when the Berry curvature is zero in the whole BZ except at the band crossing points.

\subsection{Model exhibiting a flat band with single QBCP}
A single BCP can exist alone in the BZ when the Berry curvature does not vanish.
In this section, we construct a tight-binding model realizing this possibility.
The model is defined on the honeycomb lattice.
Here, we use the same convention introduced in the previous section for the honeycomb lattice.

The tight-binding Hamiltonian is given by
\ba
H_{\rm single}(\bk) = \frac{1}{4}\bpm |g_2(\bk)|^2 & -g_1(\bk)g_2^*(\bk) \\ -g_1^*(\bk)g_2(\bk) & |g_1(\bk)|^2 \epm,
\label{eq:Htb4}
\ea
where
\bg
g_1(\bk) = e^{-\frac{i}{3}(k_1+k_2)} \left[2-(1+i)e^{ik_2}-(1-i)e^{-ik_1+ik_2}\right], \nn \\
g_2(\bk) = e^{-\frac{i}{3}(2k_1-k_2)} \left(2-e^{-ik_2}-e^{ik_1-ik_2}\right).
\eg
Note that this system is symmetric under a combined symmetry operation of mirror $M_x$ and time reversal $T$:
\ba
U(M_xT) H_{\rm single}(\bk) U(M_xT)^{-1} = H_{\rm single}(k_x,-k_y),
\ea
where $U_{M_xT}$ is given by the complex conjugation $\mc{K}$.

A wave function of flat band and the corresponding CLS are given by $\ket{\psi_{\rm flat}(\bk)} = (g_1(\bk),g_2(\bk))^T$ and
\bg
\ket{w(\b R)} = 2c^\dg_{1,\b R} -(1+i) c^\dg_{1,\b R-\b a_2} - (1-i) c^\dg_{1,\b R+\b a_1-\b a_2} \nn \\
+2c^\dg_{2,\b R} - c^\dg_{2,\b R+\b a_2} - c^\dg_{2,\b R-\b a_1+\b a_2},
\eg
respectively.

The band structure exhibits a single QBCP only at $\Gamma=(0,0)$ [See Fig.~\magenta{4}(a) in the main text].
For the BCP at $\Gamma$, $\mc{C}_{\rm BS}$ on the Bloch sphere has a diameter $\dm=1/\sqrt{2}$ as shown in Fig.~\magenta{4}(b).
Since the Berry curvature $F_{xy}(\bk)$ does not vanish, the Berry phase is not path independent.
To see this, let us consider a hexagonal path encircling the $\Gamma$, $C(L)$ where $L$ denotes the length of the side.
We calculate the Berry phase $\Phi_B[C(L)]$ along the $C(L)$ for $L\in[0,\frac{4\pi}{3}]$, as shown in Fig.~\magenta{4}(c).
Note that the limiting value of $\Phi_B[C(L)]$ for $L\rightarrow 0$ is $\sqrt{2}\pi$ as expected from the value of $\dm$, and the Berry phase along the BZ boundary $\Phi_B[C(\frac{4\pi}{3})]$ is zero.
Hence, a single BCP can exist alone in the BZ since the non-zero Berry curvature makes the Berry phase along the BZ boundary zero.


\section{Multi-band model \label{app:MB_Model}}
In this section, we consider a multi-band model beyond the two-band system.
In general, the band crossings in multi-band model can be divided into two cases depending on the degeneracy at the BCP: 
1) Two bands, which are crossing linearly or quadratically, are decoupled from other bands,
and 2) more than three bands are crossing at a point
as in the Lieb lattice.
Here, we focus on the first case parallel to the discussion in the main text.
Studying the second case would be an interesting problem for future study.

For the first case, let us consider a multi-band model in which only two bands cross each other at $\bk_0$, and other bands are isolated from these bands with a finite gap.
Then, we can divide the orthonormal eigenstates into ``crossing bands'' $\ket{\psi_n}$ $(n=1,2)$ and ``non-crossing bands'' $\ket{\psi_a}$ $(a=3,\dots,n_{\rm tot})$.
The effective Hamiltonian $\mc{H}(\b q)$ for the ``crossing bands'' near $\bk_0$ can be derived from the L{\"o}wdin perturbation theory~\cite{lowdin1951note}:
\begin{widetext}
\ba
\mc{H}(\b q)_{nm}
= H(\bk_0)_{nm}
+ \sum_{i=1}^d H_i(\bk_0)_{nm} q_i
+ \sum_{i,j=1}^d \left[ \frac{1}{2} H_{ij}(\bk_0)_{nm} + \sum_{a=3}^{n_{\rm tot}} \frac{H_i(\bk_0)_{na} H_j(\bk_0)_{am}}{E_n(\bk_0)-E_a(\bk_0)} \right] q_i q_j + O(q^3),
\ea
\end{widetext}
where $\b q=\bk-\bk_0$, $d=2$ denotes the dimensionality, and
\ba
H(\bk_0)_{\mu\nu}&=\bra{\psi_\mu(\bk_0)} H(\bk_0) \ket{\psi_\nu(\bk_0)}, \\ H_i(\bk_0)_{\mu\nu}&=\bra{\psi_\mu(\bk_0)} \der_i H(\bk_0) \ket{\psi_\nu(\bk_0)}, \\
H_{ij}(\bk_0)_{\mu\nu}&=\bra{\psi_\mu(\bk_0)} \der_i \der_j H(\bk_0) \ket{\psi_\nu(\bk_0)}.
\ea
Note that the lower indices $(n,m)$ run for the crossing bands while $a$ runs for the non-crossing bands, and the band indices ($\mu,\nu$) run over both crossing and non-crossing bands.
With the two-band effective Hamiltonian $\mc{H}_{\rm eff}(\b q)$, the BCP in the multi-band model can be analyzed in a way similar to the method applied to two-band systems in the main text.

\begin{figure}[t!]
\centering
\includegraphics[width=0.48\textwidth]{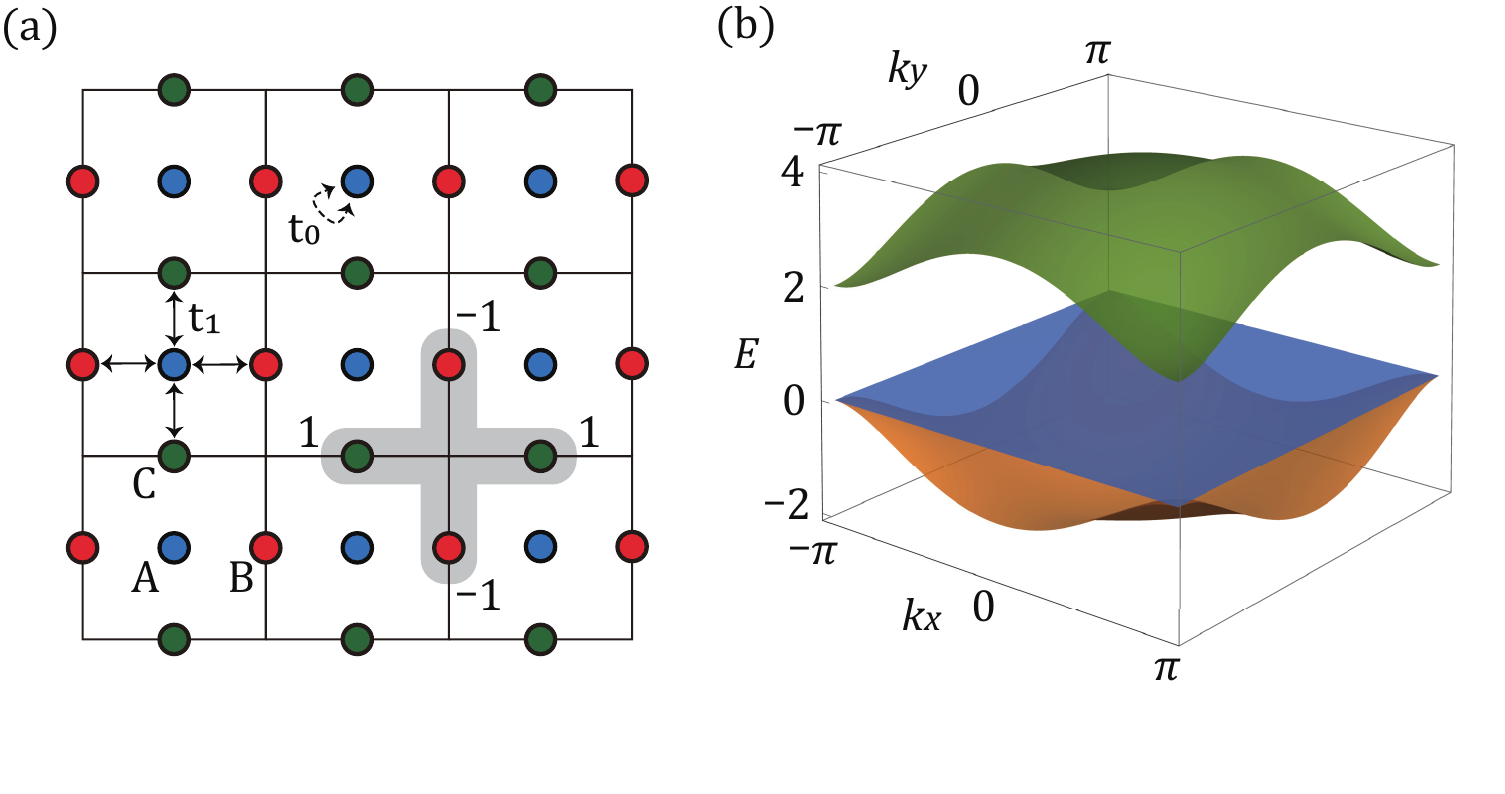}
\caption{
Lattice system and band structure of $H_{\rm Lieb}(\bk)$.
(a) The Lieb lattice is composed of three sublattice sites, $A$, $B$, and $C$.
The solid and dashed arrow denote the hopping interactions.
The shaded region illustrates a schematic description of a localized eigenstate for the flat band, $\ket{\psi_{\rm flat}(\bk)}$.
The numbers near the sublattice sites indicate the amplitudes in the eigenstate.
(b) The band structure for $H_{\rm Lieb}(\bk)$ with $(t_0,t_1)=(2.0,1.0)$.
A flat band exhibits a QBCP at $\bk=(\pi,\pi)$.
}
\label{FigS1}
\end{figure}

Now, we apply this method to a tight-binding model in the Lieb lattice.
The tight-binding Hamiltonian in momentum space is given by
\ba
H_{\rm Lieb}(\bk)= \bpm t_0 & 2t_1\cos\frac{k_x}{2} & 2t_1\cos\frac{k_y}{2} \\
2t_1\cos\frac{k_x}{2} & 0 & 0 \\
2t_1\cos\frac{k_y}{2} & 0 & 0 \epm,
\label{eq:threebandH}
\ea
where $t_0$ and $t_1$ denote the on-site energy of sublattice site $A$ and the strength of nearest-neighbor hoppings between sublattice sites $B$ and $C$, respectively [See \fig{FigS1}(a).]
This model exhibits the band structure in which the second lowest energy band is completely flat.
Also, the two lowest (highest) energy bands exhibit a quadratic band crossing at $\bk_0=(\pi,\pi)$ when $t_0$ is positive (negative).
The band structure for $t_0>0$ is shown in \fig{FigS1}(b).
First, let us consider a wave function of flat band,
\ba
\ket{\psi_{\rm flat}(\bk)} = \left( 0, -\cos\frac{k_y}{2}, \cos\frac{k_x}{2} \right)^T.
\ea
Plugging this wave function into the general definition in \eq{eq:qdis_supp}, we obtain straightforwardly that the maximum quantum distance $\dm=1$.
%

Now, we obtain the same result using the L{\"o}wdin perturbation theory.
At the BCP, the eigenstates are given by $\ket{\psi_1(\bk_0)}=(1,0,0)^T$, $\ket{\psi_2(\bk_0)}=(0,1,0)^T$, and $\ket{\psi_3(\bk_0)}=(0,0,1)^T$ with the corresponding energy eigenvalues $E_1(\bk_0)=E_2(\bk_0)=0$ and $E_3(\bk_0)=t_0$.
Accordingly, the effective Hamiltonian for $\ket{\psi_{1,2}(\bk)}$ are given by
\ba
\mc{H}_{\rm Lieb}(\b q)
= -\frac{t_1^2}{t_0} \bpm q_x^2 & q_x q_y \\ q_x q_y & q_y^2 \epm,
\label{eq:effH_Lieb}
\ea
where $\b q=\bk -\bk_0$.
Diagonalizing $\mc{H}_{\rm Lieb}(\b q)$, we obtain the eigenstates $\ket{\psi_{\rm flat}(\b q)}=(-q_y,q_x)^T$ and $\ket{\psi_{\rm quad}(\b q)}=(q_x,q_y)^T$ with the corresponding energy eigenvalues $E_{\rm flat}(\b q)=0$ and $E_{\rm quad}(\b q)=-\frac{t_1^2}{t_0} q^2$.
Consistent with the discussion in the main text, as the trajectory on the Bloch sphere forms a great circle, one can show that the BCP in this system exhibits $\dm=1$ and the zero Berry phase.
Thus, the analysis based on the full Hamiltonian $H_{\rm Lieb}(\bk)$ in \eq{eq:threebandH} and one based on the effective Hamiltonian $\mc{H}_{\rm Lieb}(\b q)$ in \eq{eq:effH_Lieb} give the same result.

\end{document}